\documentclass{emulateapj}
\usepackage{graphicx}

\def\ltsima{$\; \buildrel < \over \sim \;$}
\def\simlt{\lower.5ex\hbox{\ltsima}}
\def\gtsima{$\; \buildrel > \over \sim \;$}
\def\simgt{\lower.5ex\hbox{\gtsima}}
\def\kms{{\rm\,km\,s^{-1}}}

\def\kpc{{\rm\,kpc}}

\def\msun{{\rm\,M_\odot}}
\def\msunyr{{\rm\,M_\odot/yr}}
\def\lsun{{\rm\,L_\odot}}

\def\pc{{\rm\,pc}}
\def\cm{{\rm\,cm}}

\newcommand{\fmmm}[1]{\mbox{$#1$}}
\newcommand{\scnd}{\mbox{\fmmm{''}\hskip-0.3em .}}
\newcommand{\scnp}{\mbox{\fmmm{''}}}

\def\AA{$\; \buildrel \circ \over {\rm A}$}

\def\deg{^\circ}

\def\s{\ifmmode \widetilde \else \~\fi}
\def\={\overline}

\def\spose#1{\hbox to 0pt{#1\hss}}

\def\etal{{\it et al.\ }}
\def\cf{{\it cf.\ }}
\def\eg{{ e.g.,\ }}

\def\lta{\mathrel{\spose{\lower 3pt\hbox{$\mathchar"218$}}
     \raise 2.0pt\hbox{$\mathchar"13C$}}}
\def\gta{\mathrel{\spose{\lower 3pt\hbox{$\mathchar"218$}}
     \raise 2.0pt\hbox{$\mathchar"13E$}}}
\def\Dt{\spose{\raise 1.5ex\hbox{\hskip3pt$\mathchar"201$}}}    
\def\dt{\spose{\raise 1.0ex\hbox{\hskip2pt$\mathchar"201$}}}    

\def\dotsfill{\leaders\hbox to 1em{\hss.\hss}\hfill}

\def\Gyr{{\rm\,Gyr}}
\def\Myr{{\rm\,Myr}}

\begin{document}

\title{On the accretion origin of a vast extended stellar disk around the
Andromeda galaxy}
\author{R. Ibata\altaffilmark{1}, 
S. Chapman\altaffilmark{2}, 
A. M. N. Ferguson\altaffilmark{3}, 
G. Lewis\altaffilmark{4}, 
M. Irwin\altaffilmark{5}, 
N. Tanvir\altaffilmark{6}}
\altaffiltext{1}{
Observatoire de Strasbourg, 11, rue de l'Universit\'e, F-67000, Strasbourg,
France} 
\altaffiltext{2}{California Institute of Technology, Pasadena, CA, 91125}
\altaffiltext{3}{Institute for Astronomy, University of Edinburgh,
Royal Observatory, Blackford Hill, Edinburgh, UK EH9 3HJ}
\altaffiltext{4}{Institute of Astronomy, School of Physics, A29, University of Sydney, NSW
2006, Australia}
\altaffiltext{5}{Institute of Astronomy, Madingley Road, Cambridge, CB3 0HA, U.K.}
\altaffiltext{6}{Physical Sciences, Univ. of Hertfordshire, Hatfield, AL10 9AB, UK}

\begin{abstract}
We present  the discovery of an  inhomogenous, low-surface brightness,
extended disk-like  structure around the Andromeda  galaxy (M31) based
on  a  large kinematic  survey  of more  than  $2800$  stars with  the
Keck/DEIMOS  multi-object spectrograph.   The stellar  structure spans
radii from  $15\kpc$ out to $\sim  40\kpc$, with detections  out to $R
\sim  70\kpc$.  The  constituent stars  lag the  expected  velocity of
circular orbits in the plane of the M31 disk by $\sim 40\kms$ and have
a velocity dispersion of $\sim 30\kms$.  The color range on the upper
RGB shows a large spread indicative of a population with a significant
range  of  metallicity.   The  mean  metallicity  of  the  population,
measured from Ca~II  equivalent widths, is ${[Fe/H] =  -0.9 \pm 0.2}$.
The morphology of the structure is irregular at large radii, and shows
a wealth of substructures which  must be transitory in nature, and are
almost  certainly tidal  debris. The  presence of  these substructures
indicates  that  the global  entity  was  formed  by accretion.   This
extended disk follows  smoothly on from the central  parts of M31 disk
out to $\sim  40\kpc$ with an exponential density  law of scale-length
of $5.1  \pm 0.1 \kpc$, which is  similar to that of  the bright inner
disk.   However,  the   population  possesses  similar  kinematic  and
abundance properties  over the entire  region where it is  detected in
the survey.  We estimate that the structure accounts for approximately
10\%  of the  total luminosity  of the  M31 disk,  and given  the huge
scale, contains $\sim 30$\% of  the total disk angular momentum.  This
finding indicates that at least some galactic stellar disks are vastly
larger than previously thought and are formed, at least in their outer
regions, primarily by accretion.
\end{abstract}

\keywords{galaxies: individual (M31) --- galaxies: structure --- 
galaxies: evolution --- Local Group}

\section{Introduction}

Within the framework of hierarchical structure formation, large spiral
galaxies like the  Milky Way or Andromeda (M31)  arose from the merger
of  many small galaxies  and protogalaxies  which began  coalescing at
high  redshift.   During  this  process,  considerable  dynamical  and
chemical evolution took place, as the primordial galaxy fragments were
assimilated,  stellar populations  evolved and  fed enriched  gas back
into  the interstellar  medium, and  fresh gas  was accreted  onto the
galaxy.  The  disk, the defining component of  spiral galaxies, formed
through dissipative  processes, growing from the inside  out mainly by
the accretion of  gas with high angular momentum  with probably only a
small component of accreted stars \citep{abadi}.

\begin{figure*}
\begin{center}
\includegraphics[angle=0,width=12cm]{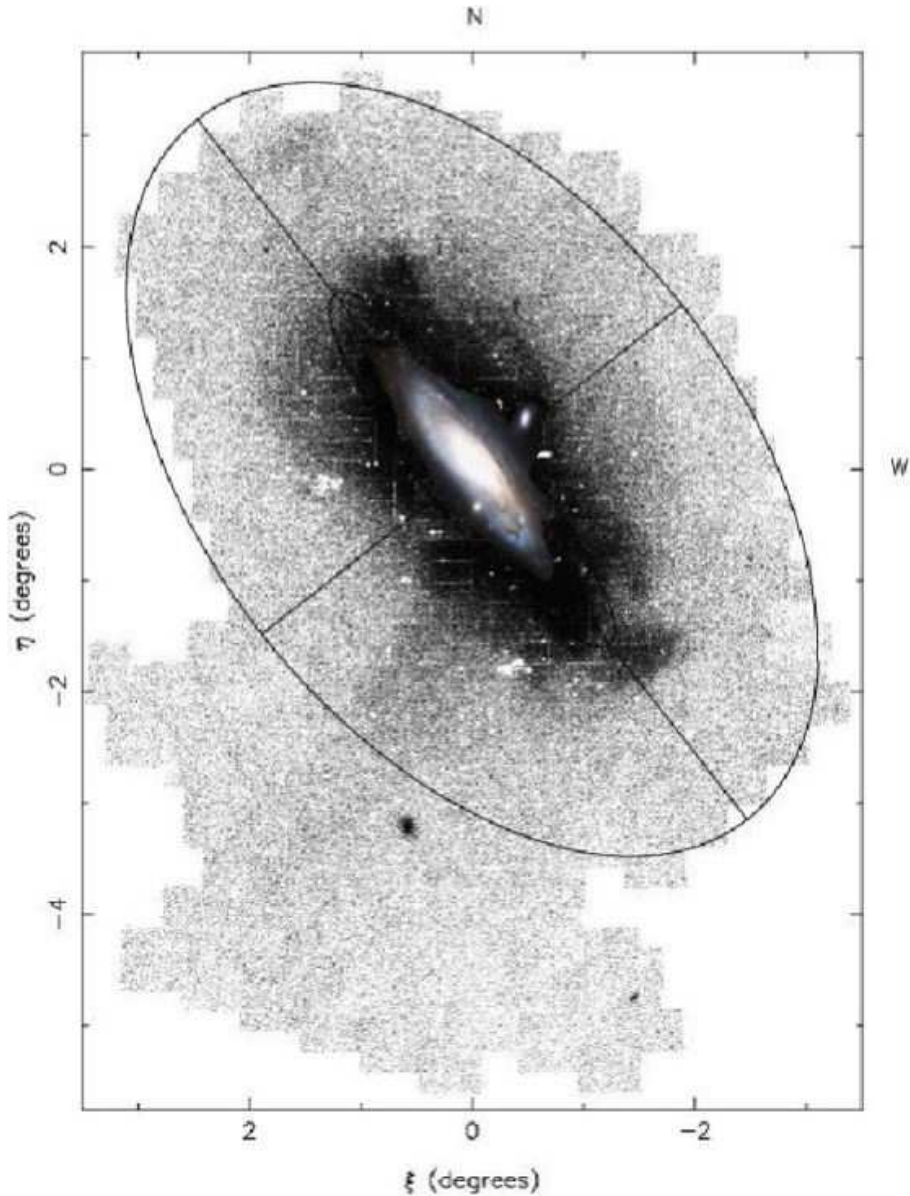}
\end{center}
\caption{The  current coverage of  our large  panoramic survey  of M31
with the  INT camera, in standard coordinates  $(\xi,\eta)$. The outer
ellipse  shows a  segment of  a $55\kpc$  radius ellipse  flattened to
$c/a=0.6$, and  the major and  minor axis are indicated  with straight
lines out to this ellipse. A  DSS image of the central galaxy has been
inserted  to  scale. The  map  shows  the  distribution of  RGB  stars
throughout the halo.  The most striking structure in the  map is a vast
flattened inner  structure, of major  axis extent $\sim 4$  degrees, a
messy inhomogenous  entity which envelops the familiar  bulge and disk
of M31.}
\end{figure*}

The structure of  disks is generally exponential in  both their radial
and  vertical directions.   A striking  feature of  this  component of
galaxies  is the  tendency for  it to  be truncated  radially  at 3--4
scale-lengths \citep{kruit,  pohlen}.  This distance  would correspond
to  $\sim 12  \kpc$  in  the Milky  Way  and $\sim  20  \kpc$ in  M31.
Although  we still lack  a solid  understanding of  how and  why these
truncations  occur, evidence  suggests an  abrupt change  in  the star
formation efficiency across this  boundary, perhaps related to a shift
in  the  dominant star  formation  mode \citep{kennicutt,  ferguson98,
schaye}. The  fact that disk  stars are detected beyond  the predicted
truncation radii in the Milky Way and M31, as well as in several lower
mass galaxies  (e.g. NGC~300, \citealt{bland-hawthorn};  M33, Ferguson
et al  2005, in prep),  is intriguing.  Did  these stars form  {\it in
situ}?  If  so, did they  form a long  time ago during the  very early
stages of disk evolution, or  have they formed much more recently from
newly-acquired gas?  On  the other hand, could they  have been tidally
stripped from infalling satellites on close to coplanar orbits?

In  order to  address  these questions  and  examine galaxy  formation
models in detail,  our group has undertaken a  deep panoramic study of
the Andromeda galaxy,  which when set in contrast  with the Milky Way,
was to  serve as  a test case  against which  formation simulations
could be compared.  Studies of  the Milky Way are naturally blessed with
more  photons that  those of  more distant  systems, and  by resolving
individual stars and measuring their kinematic and chemical properties
one can study the formation of our  Galaxy in a detail and with a mass
and  spatial  resolution  that  can  only  be  dreamed  of  elsewhere.
However,  thanks  to  advances  in  instrumentation, it  is  now  also
possible to  measure the kinematics and chemistry  of individual stars
in galaxies of  the Local Group.  One of the  major benefits that this
affords us is a  global view of a galaxy, freeing us  from much of the
projection  uncertainties  that  our  position within  the  Milky  Way
plagues us with, and yet  retaining the fine detail that spectroscopic
studies of resolved stellar populations provide.

Located at a  distance of $785\pm25\kpc$ ($(m -  M)_0 = 24.47\pm0.07$;
\citealt{mcconnachie05}), Andromeda is the closest giant spiral galaxy
to our  own, and the  only other giant  galaxy in the Local  Group. As
such it is  often considered to be the ``sister''  galaxy to the Milky
Way.  Recent studies suggest that the two galaxies do indeed have very
similar  total masses (including  the dark  matter, \citealt{evans00b,
ibata04}), yet there are significant differences between them.  M31 is
slightly more  luminous than the Milky  Way, it has  a higher rotation
curve, and a bulge with  higher velocity dispersion. M31 possesses and
a globular cluster system with $\sim 500$ members, approximately three
times more numerous  than that of the Milky  Way, and contains several
members of a type of  luminous extended cluster not present around the
Galaxy  \citep{huxor05}.  The  disk  of Andromeda  is  also much  more
extensive, with  a scale-length  of $5.9 \pm  0.3 \kpc$  (R-band value
corrected for a distance of $780\kpc$, \citealt{walterbos88}) compared
to $2.3\pm0.1$ for the Milky Way \citep{ruphy}; but which is currently
forming  stars at  a lower  rate than  the  Galaxy \citep{avila-reese,
walterbos94}.    Though  possibly   the   consequence  of   low-number
statistics, it is tempting to  attribute significance to the fact that
Andromeda has  four dwarf elliptical galaxies  (M32, NGC~205, NGC~147,
NGC~185) among  its entourage of satellites and  no star-forming dwarf
irregulars, whereas  the Milky Way has  none of the former  but two of
the latter.  However, it is perhaps in their halo populations that the
differences  between  the  two  galaxies  are most  curious  and  most
interesting.  It  has been known  for many years  that M31 
does not possess the ``text-book'' stellar halo model developed
from observations of the Milky Way.  Minor  axis halo
fields have  shown startlingly  high metallicities (${\rm  [Fe/H] \sim
-0.5}$) and an $R^{1/4}$-law density profile \citep{durrell04}, unlike
the   $R^{-3.5}$   behavior   of   the   metal-poor   Galactic   halo
\citep{chiba00}.  Nevertheless, recent studies  of M31 confirm that an
old,  metal-poor  halo   lurks  beneath  \citep{holland96,  durrell01,
reitzel02,  brown}.  Both  the  de Vaucouleurs  profile  and the  high
metallicity are suggestive of an  active merger history at the time of
halo (or bulge) formation.

The photometric  dataset that forms  the foundation of this  study was
obtained  with  the  Wide  Field  Camera  at  the  2.5m  Isaac  Newton
Telescope;  the  survey  and  preliminary results  were  presented  in
\citet{ibata01b}  and \citet{ferguson02},  though  the full  completed
survey will be discussed in  detail in a forthcoming article (Irwin et
al., in prep).  The density of Red Giant Branch (RGB) stars throughout
the  survey  region,  which  now  covers $\sim  40$~sq.   degrees,  is
reproduced in  Figure~1, onto which we have  superimposed the familiar
Digital  Sky Survey  image.  The  solid  angle covered  by the  survey
corresponds  to a projected  area of  $\sim 7400  \kpc^2$  at the
distance of M31.  As shown in the earlier publications in this series,
the  INT  survey revealed  a  wealth  of  substructure throughout  the
``halo'' of M31.   However, the most prominent of  the details that we
uncovered was the presence  of a gigantic extended flattened structure
around  M31, spanning  up to  $\sim  40$~kpc from  the galaxy  centre,
almost twice the maximum extent of  the bright inner disk.  Due to the
expectations from galaxy formation  simulations with cold dark matter,
our initial interpretation of  this structure favored the possibility
that it was a  flattened inner halo \citep{katz91, summers, steinmetz}
with  significant  spatial  sub-structure  \citep{klypin99,  moore99}.
However,  the proximity  of  M31 has  allowed  us to  follow up  these
detections with  deep Keck/DEIMOS spectroscopic  observations (initial
results    of   the   spectroscopic    survey   were    presented   in
\citealt{ibata04}).   It is  the radial  velocities  and metallicities
derived from these deep spectroscopic data that we present and analyze
in this contribution.

The layout of  this paper is as follows. In \S2\  we first discuss the
global geometry of  M31, the inclination angle of the  disk and fit an
exponential function to the major  axis photometry. The spectroscopic observations are
presented  in \S3, and  the resulting  kinematics in  \S4. In  \S5\ we
examine  several fields  in more  detail, presenting  also  their mean
metallicity and  fitting global trends. These data  are interpreted in
\S6, where we  present several possibilities for the  formation of the
structure, and we finish in \S7\ with the conclusions of the study.

\section{Simple global structural model of M31}

Before  describing  the observations  and  analysis  of the  kinematic
dataset,  it  is  useful  to  discuss  the  large-scale  structure  of
M31. Andromeda is  observed highly inclined to the  line of sight with
$i =  77\deg$ \citep{walterbos88}, and  it is seen  on the sky  with a
position angle  of $38\deg$ East of  North \citep{walterbos87}. Within
an angular  distance of $90\arcmin$  ($20.4\kpc$), \citet{walterbos88}
find a  good fit with the  sum of a  bulge model together with  a disk
model  of  R-band scale-length  $5.9  \pm  0.3  \kpc$ (corrected for
a distance of $780\kpc$). 

To compare  these findings  to our INT  survey, we measure  the source
density in a color-magnitude box designed to select the brightest RGB
stars  1.5  magnitudes  below   the  RGB  tip.   Sources  with  I-band
magnitudes between ${\rm  20.0 < i_0 < 21.5}$ and  colors ${\rm 1.0 <
(V-i)_{0} < 4.0}$ are selected.  (The \citealt{schlegel} dust maps are
used  throughout   this  work  to  correct  the   INT  photometry  for
extinction).   We  correct  the  star-counts for  crowding  using  the
approach of \citet{irwin84}, and  subtract a small background count as
determined from fields at the edge of the survey.

The  de-projected positions of  the RGB  stars are  binned in  a polar
coordinate grid with $5\deg$ azimuthal bins and $0.5\kpc$ radial bins.
The azimuthal counts  at a given radius are  then medianed together in
groups of 5 (i.e. covering $25\deg$), and we step around the circle in
intervals  of  $15\deg$. The  upper  panel  of  Figure~2 shows  the  4
profiles within a de-projected angle of $\pm 35\deg$ of the major axis
(i.e.   $\pm 7.9\deg$  projected),  assuming an  inclination  of $i  =
77\deg$, with the additional thick-lined profile showing the median of
all 14 bins of $5\deg$ width.  Positive  radius means that the medianed area is
on  the   North-eastern  major  axis,  and  negative   radius  on  the
South-western major axis.  The abscissa $R$ is a de-projected radius,
that  assumes the  stated  inclination.  The  artificial  drop in  the
source counts inside of $R \sim 18\kpc$ is due to crowding.

\begin{figure}
\begin{center}
\includegraphics[angle=0,width=\hsize]{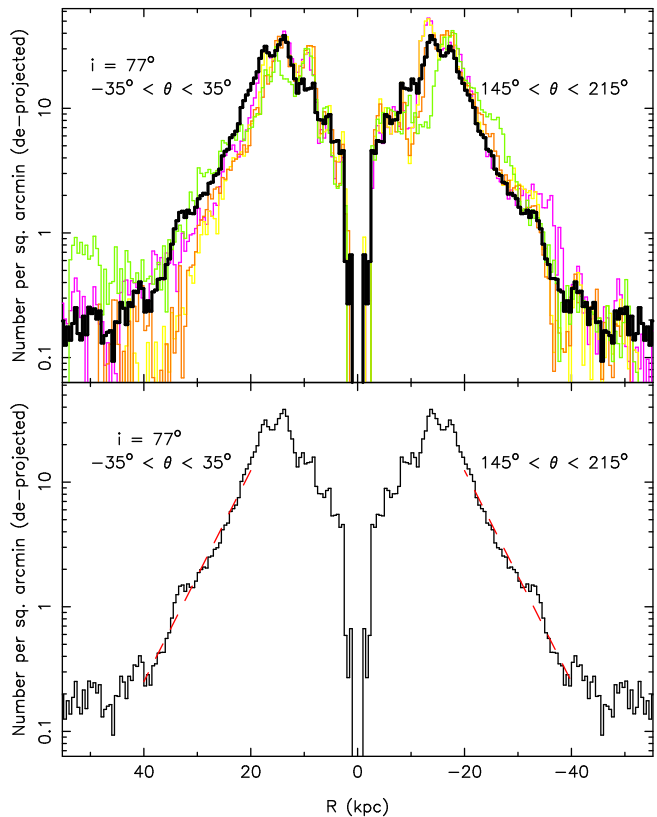}
\end{center}
\caption{The top panel  shows the density profiles close  to the major
axis.   Positive  abscissa  values  correspond to  de-projected  radii
towards the North-eastern side of  M31. The thin lines are the density
profiles medianed over 5 $5\deg$-de-projected azimuthal bins, assuming
a  galaxy  inclination  of  $i  =  77\deg$.  The  4  profiles  to  the
North-east cover the de-projected angles $-35\deg < \theta < -10\deg$,
$-20\deg  < \theta <  5\deg$, $-5\deg  < \theta  < 20\deg$,  $10\deg <
\theta  <  35\deg$,  with  the  4 profiles  to  the  South-west  being
identical cuts rotated by $180\deg$.  The thick black line corresponds
to the  median of all  14 azimuthal bins  between $-35\deg <  \theta <
35\deg$.  The  lower panel  shows an exponential  fit to  the medianed
profile between  $20 < R <  40\kpc$, with a scale-length  $5.1 \pm 0.1
\kpc$.}
\end{figure}

Evidently the major axis profile is close to exponential; it is fit in
the lower  panel with a model  (dashed line) of  scale-length $5.1 \pm
0.1 \kpc$. This scale-length is similar  to that of the inner disk, so
it appears that the large-scale properties of the outer disk structure
follow on smoothly from those of the inner disk.

The  upper panel  of Figure~3  shows  the profiles  for all  azimuthal
bins. The large discrepancies  between the different profiles are very
obvious, with the less  steep profiles corresponding to azimuthal bins
towards the minor axis. The  higher number densities on the minor axis
are largely  due to the  huge bulge component  in M31. Indeed,  in the
bulge-disk decomposition  of \citet{walterbos88}, the  bulge dominates
over the  disk at all radii on  the minor axis. However,  part of this
variation may also come from a change in the inclination of the galaxy
at large radius.  To explore this possibility, we  fitted the value of
$i$ by assuming that the most  likely inclination is that that has the
lowest root  mean square  dispersion about the  median profile  in the
distance interval  $20 < R  < 40\kpc$. The corresponding  profiles are
shown  on the  lower panel  of  Figure~3, together  with the  medianed
profile  (thick   black  line).   The  fitted   inclination  angle  is
$i = 64.7\deg$.

\begin{figure}
\begin{center}
\includegraphics[angle=0,width=\hsize]{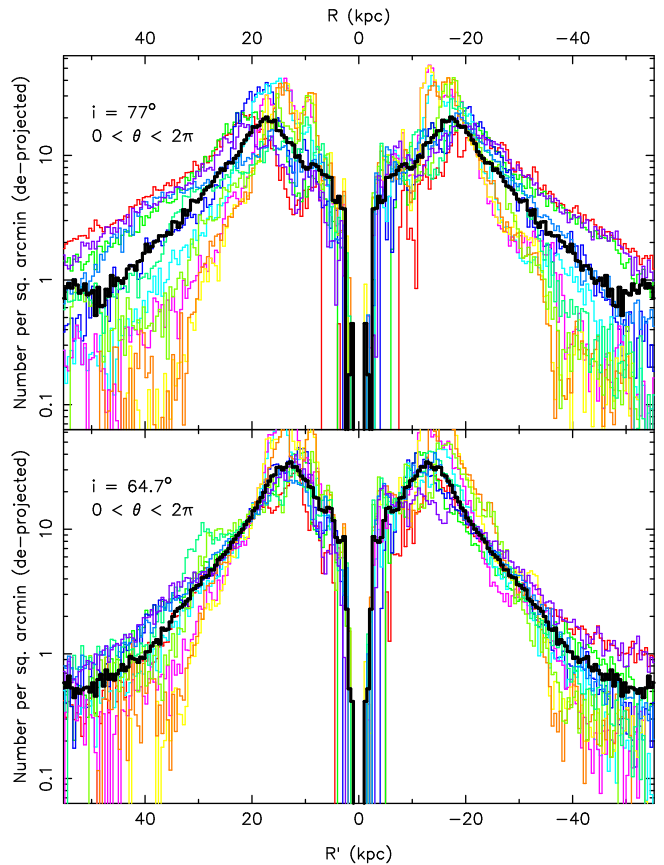}
\end{center}
\caption{The top  panel shows  medianed density profiles  at different
position  angles  around  M31,  offset  at  $15\deg$  from  each other,
assuming a  galaxy inclination of $i  = 77\deg$. As  in Figure~2, each
profile is constructed  from a median of 5  consecutive azimuthal bins
of width $5\deg$.   The thick black line corresponds  to the median of
all azimuthal bins. The lower  panel shows similar profiles, but for a
an  inclination  angle of  $i  =  64.7\deg$,  fitted to  minimize  the
dispersion around  the median profile  of all azimuthal bins  over the
interval $20 < R^\prime < 40\kpc$.}
\end{figure}

With this  smaller inclination, the  surface density profiles  are now
much more similar, yet there are still variations between the profiles
of up to a  factor of $\sim 2$--$3$ at a given  radius, due in part to
the presence of obvious ``lumps''.  Fitting the median profile between
$20 <  R' < 40\kpc$ with  an exponential-law yields  a scale-length of
$7.8 \pm  0.07\kpc$.  This larger  scale-length value compared  to the
previous  major axis  exponential fit  demonstrates a  failure  in our
simple tilted disk model. A  more complex model with a bulge component
is required  to fit the central  regions of the  galaxy more reliably,
yet  this cannot be  constructed from  our INT  data which  are highly
incomplete inside of $R \sim 15\kpc$.

In  the  following  discussion,  we  will  present  results  based  on
inclination values of $i = 77\deg$  and $i = 64.7\deg$; this will help
clarify the uncertainty due to a  possible tilt of the outer disk with
respect  to the  inner  disk.  De-projected  quantities,  such as  the
de-projected radius $R$ or  the velocity lag $v_{lag}$ (defined below)
will implicitly assume an inclination of $i = 77\deg$; primed versions
of these  variables (e.g.  $R^\prime$,  $v^\prime_{lag}$) will instead
correspond to the lower inclination case of $i = 64.7\deg$.

\section{Observations}

During  eight Keck  nights ($\sim75$\%  useful time)  in  Sept.\ 2002,
2003,  and 2004, we  undertook a  DEIMOS program  to spectroscopically
follow-up  the   detected  substructures  in  M31   to  measure  their
kinematics,  and  to  begin  a  complete  kinematic  portrait  of  M31
targeting  systematically disk  and halo  fields.  The  observing runs
have  employed the  standard  high-res {\it  DEEP2} slitmask  approach
(e.g., \citealt{davis03}) covering the wavelength range 6400--9000\AA,
with  a spectral  resolution of  $\sim$0.6\AA.  The  observations have
also  pioneered  two  new  approaches  with  DEIMOS:  a  band-limiting
$\sim300$\AA\ Ca~II triplet  filter to multiplex  $\sim 4\scnp$
slitlets in the  spectral direction yielding as many  as 800 slits per
mask,  and a  `fibre-hole' $0\scnd 7$  slitlet approach  packing $\sim
600$ holes per mask. This `fibre-hole' approach adopted in Sept.\ 2004
proved to  be very successful,  giving Poisson-limited sky-subtraction
(down  to i=21.5)  by assigning  holes  to monitor  the sky  spectrum;
longer slitlets  simply waste valuable  CCD real-estate in  regions of
high target density.  The  fiber-hole diameters were milled at $0\scnd
7$ to match the median seeing.  The dataset presented in this paper is
the sum of all the  `standard mask' spectra plus `fibre-holes' spectra
obtained  up to  September  2004  that do  not  use the  band-limiting
filter. A total of 28 fields were observed in this way.  The layout of
these  spectroscopic fields  is displayed  in Figure~4:  the  5 fields
D1--D5 probe the  M31 disk, the 16 fields F1--F16  probe both the halo
and what  we considered to  be ``halo'' overdensities, while  7 fields
were  assigned  to  investigate  the  dynamics  of  the  giant  stream
\citep{ibata01b} which is seen clearly in Figure~1 as an elongated
structure to the South-east of M31.

\begin{figure}
\begin{center}
\includegraphics[angle=270,width=\hsize]{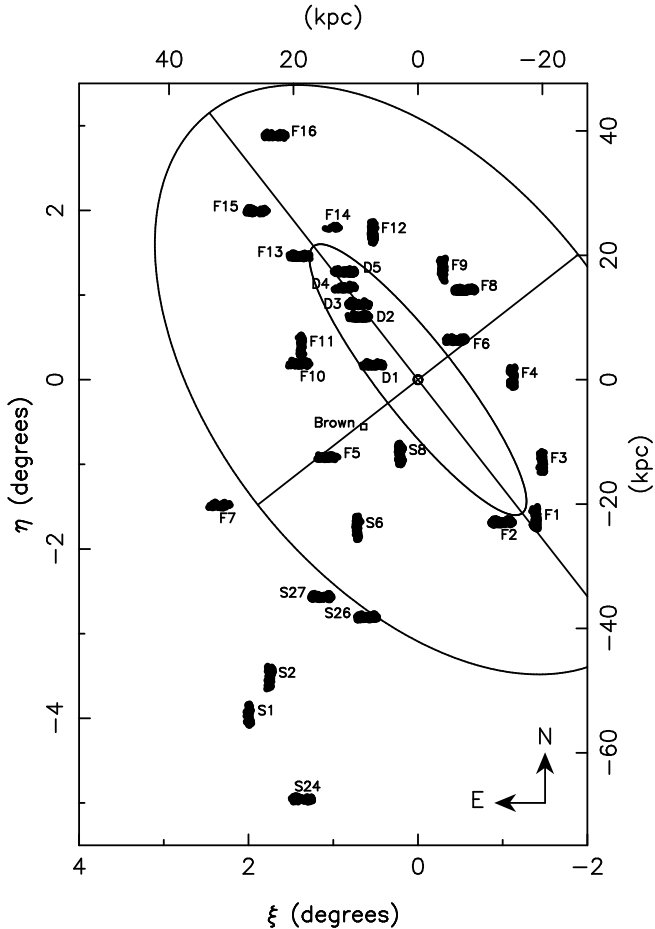}
\end{center}
\caption{The  distribution of  Keck/DEIMOS fields  in relation  to the
survey region depicted in Figure~1. For clarity, we have divided these
fields into three groups. The  fields in the giant stream are prefixed
by  the   letter  S,  and  we   keep  the  same  field   names  as  in
\citet{ibata04}, with the addition of  fields S24, S26, and S27. Those
fields  located within the  inner ellipse  are denoted  D1 --  D5. The
remaining  fields are  named  F1  -- F16.   The  small square  labeled
``Brown'' on  the South-eastern minor  axis marks the location  of the
deep  photometric  study  of  \citet{brown}, whose  relevance  to  the
present work we discuss in \S6.}
\end{figure}

For  each  of these  fields  except  the  stream fields,  the  highest
priority  targets were selected  for observation  from the  INT survey
within  a color-magnitude  box with  I-band magnitudes  between ${\rm
20.5 < i < 22.0}$ (without reddening correction) and colors ${\rm 1.0
< (V-i)_{0} < 4.0}$.  This broad  selection was designed to pick out a
wide range  of red-giant branch stars, containing  both metal-poor and
metal-rich  populations (further  details  will be  provided in  Irwin
\etal\  2005, in  prep).   Other non-saturated  targets brighter  than
${\rm i = 22}$ were also chosen by an automated selection algorithm at
lower  priority  to  fill  in  available  space  on  the  spectrograph
detector.

Each  mask was observed  for typically  three integrations  of 20\,min
each.  The spectroscopic images  were processed and combined using the
pipeline software  developed by  our group.  This  software debiasses,
performs    a   flat-field,   extracts,    wavelength-calibrates   and
sky-subtracts  the spectra.   The radial  velocities were  measured by
cross-correlating against a model  template with Gaussian functions at
the rest-frame  wavelength positions of  the Ca~II triplet  lines (the
technique     used    is    similar     to    that     discussed    by
\citealt{wilkinson}). An  estimate of the  radial velocity uncertainty
is  then obtained  for each  star from  the dispersion  of  the radial
velocity measured from the  three Ca~II triplet lines separately.  The
accuracy  of these  data  are very  good  for such  faint stars,  with
typical uncertainties of $5$--$10\kms$.   The sample presented in this
paper consists  of 2834 stars  with velocity uncertainties  $< 20\kms$
that fall into the above RGB color-magnitude selection window.

\section{Results}

The velocity distribution of all the targeted stars with ${\rm (V-i)_0
  > 1.0}$  is   displayed  on  the   top  panel  of   Figure~5.   This
distribution is a mixture of  bona-fide M31 stars plus a contamination
from  Galactic  stars.  We  have  superimposed  the expected  Galactic
contribution over a $0.25\deg$ area  in the direction of M31 according
to  the Besan{\c  c}on Galactic  starcounts model  \citep{robin}, onto
which we  have applied the color-magnitude selection  window used for
spectroscopic target selection.  Since we have not observed every star
in each  field, and since the ratio  of M31 to Milky  Way stars varies
dramatically  from the  inner fields  to the  outer fields,  we cannot
easily scale the model  predictions.  For the illustrative purposes of
Figure~5, we  have chosen a  normalization to match the  Besan{\c c}on
model to our observed distribution for velocities higher than the peak
in the model. We will return to the issue of Galactic contamination in
our analysis of  field F16 (Figure~24), placing an  upper limit on the
model normalisation, which shows that Galactic contamination is low at
Heliocentric radial  velocities below $-100\kms$.  It is  clear from a
visual comparison of  the distributions in Figure~5 that  we observe a
large  number  of  stars  in  significant  excess  over  the  expected
distribution of contaminants.  The  apparent mismatch between the peak
in the observed distribution and the Besan{\c c}on model is due to the
large  number of  stars  observed in  the  disk fields  D1--D5 on  the
North-eastern side of Andromeda,  which combine in this summed-up plot
to give a broader peak centered  on $-60\kms$. If we instead sum up the
11 fields on the South-western side  of M31 (fields F1--F4 and all the
stream fields), the rotation of  the M31 disk separates Galactic stars
from  the   majority  of  stars  in  Andromeda,   allowing  a  cleaner
discrimination. This is shown on  the bottom panel of Figure~5.  These
considerations   show  that   by  selecting   stars   with  velocities
$<-100\kms$  we can drastically  minimize the  Galactic contamination,
incurring only the  penalty of including a small  fraction of Galactic
halo stars into our sample;  furthermore these Galactic halo stars are
expected to have a relatively flat distribution in radial velocity.

\begin{figure}
\begin{center}
\includegraphics[angle=0,width=\hsize]{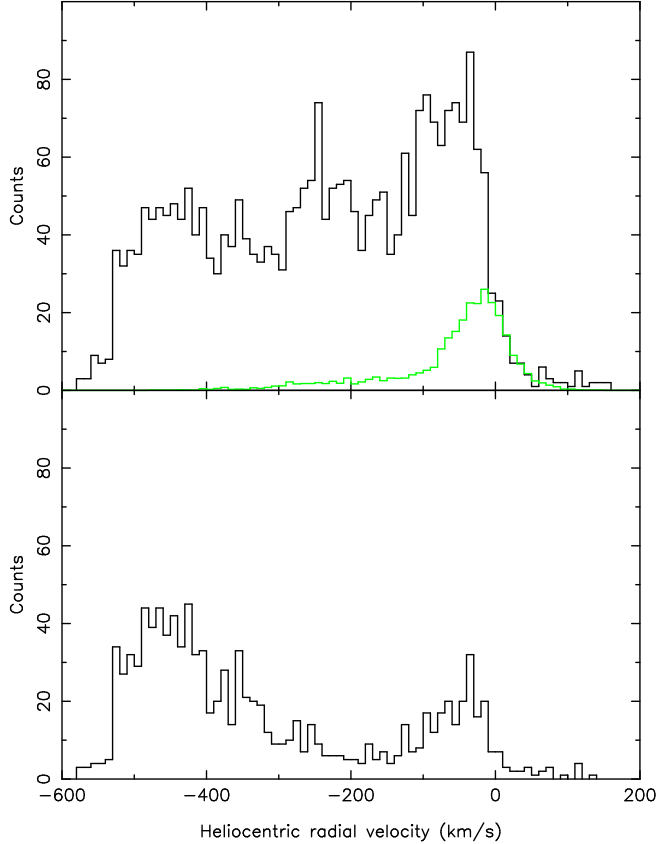}
\end{center}
\caption{The top panel shows the velocity distribution of the 2834 RGB
star  candidates  in  all  fields  shown in  Figure~4  that  pass  our
color-magnitude  and  spectrum  quality  criteria.   The  green-lined
histogram shows the prediction of the Besan{\c c}on Galactic model for
an  area  of  240~sq.~arcmin  within  our  color-magnitude  selection
window. With this normalization, the model fits the high-velocity edge
of  the observed  velocity distribution.  The bottom  panel  shows the
summed distribution of stars in  the fields F1--F4 plus all the stream
fields. }
\end{figure}

Due to the wide coverage of  the spectroscopic fields, it is useful to
compare  the observed  velocity of  a  star with  the expected  radial
velocity of  a population  moving in circular  orbits. We  construct a
simple  model assuming  the geometrical  model discussed  in  \S2.  We
adopt  the circular  velocity curve  compilation  of \citet{klypin02},
which joins CO measurements from \citet{loinard} for $R < 10\kpc$ with
\ion{H}{1} measurements by \citet{brinks}  at larger radii; these data
are reproduced in  Figure~6, along with our spline  fit which is shown
adjusted  for  the  small  inclination  correction.  The  dip  in  the
rotation curve near  $5\kpc$ is of course an  artifact of non-circular
motions  in the  gas  at that  location,  and is  unimportant for  the
present  work  since our  fields  all  lie  beyond that  radius.   The
corresponding map  of the projected  rotation velocity over  the inner
region of our survey is shown in Figure~7.

\begin{figure}
\begin{center}
\includegraphics[angle=270,width=\hsize]{M31_extendeddisk.fig06.ps}
\end{center}
\caption{The rotation curve  of M31. The points show  a rotation curve
compilation  by \citet{klypin02},  which is  derived from  CO rotation
velocities measurements \citep{loinard} for $R < 10\kpc$, and from the
rotation  velocity  of   the  \ion{H}{1}  \citep{brinks}  beyond  that
radius. The  dotted line shows a  spline fit to these  data, while the
continuous line indicates the  correction for the $77\deg$ inclination
of the inner disk of this galaxy.}
\end{figure}

\begin{figure}
\begin{center}
\includegraphics[bb= 80 130 550 600, clip, angle=270,width=\hsize]{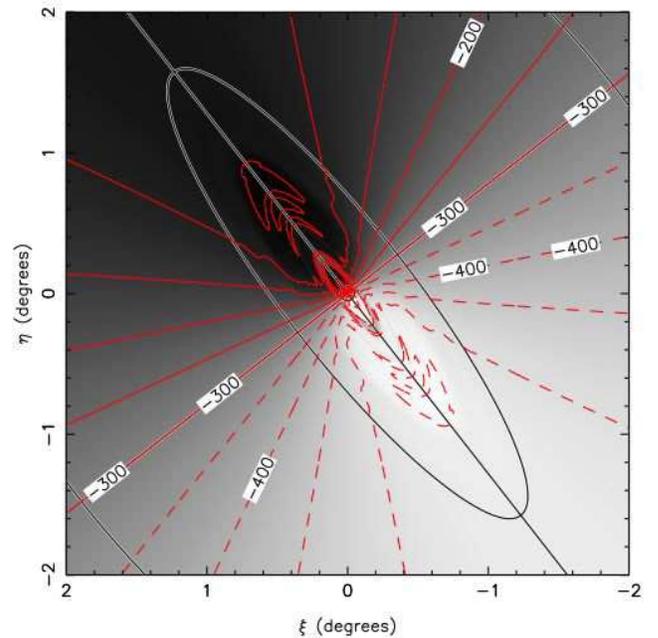}
\end{center}
\caption{A simple model for the expected velocity of stars on circular
orbits as  a function of position  around M31, made  by projecting the
spline  fit  of  Figure~6  using  the geometrical  model  proposed  in
\S2. For  comparison with  Figures 1 and  4, we have  superimposed the
inner ellipse, and  a section of the outer  ellipse displayed in those
figures onto this diagram.  The  radial velocity contours are shown at
intervals of $50\kms$.}
\end{figure}

In reality, the  M31 disk has finite thickness, and  due to the highly
inclined orientation  that it presents  to us, lines of  sight through
the disk  will probe  a substantial  depth. This is  a concern  as the
simple circular velocity model  of Figure~7 could be misleading, since
stars at a  given $(\xi,\eta)$ on the sky can  be located at different
radii in  M31, and will also  present different projections  to us. In
Figure~8  we   show  the  velocity   uncertainty  caused  by   a  disk
scale-height of $0.35\kpc$, a  value comparable to the scale-height of
the  old disk of  the Milky  Way \citep{gilmore83}.   For most  of the
survey area the additional  uncertainty to the circular velocity model
due to a disk of that  thickness is less than $10\kms$, though certain
regions within  the inner ellipse  the uncertainty is $>  30\kms$.  We
will  bear this consideration  in mind  when discussing  the kinematic
properties of the disk.

\begin{figure}
\begin{center}
\includegraphics[bb= 80 130 550 600, clip, angle=270,width=\hsize]{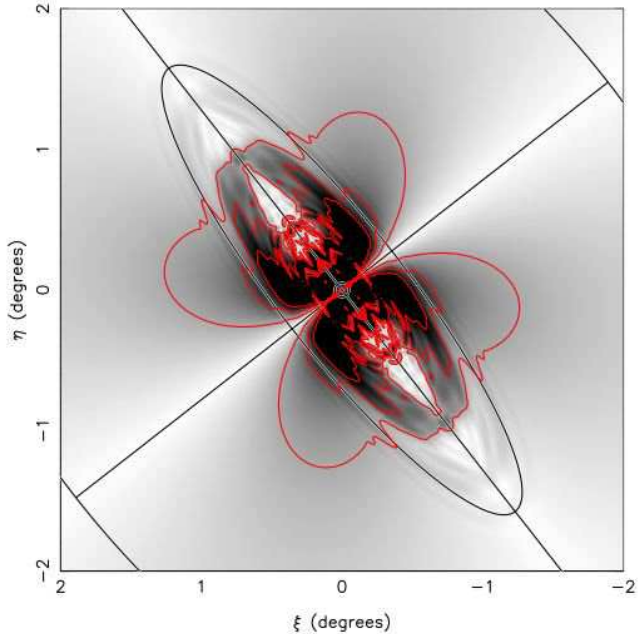}
\end{center}
\caption{Since  the Andromeda  galaxy  is observed  close to  edge-on,
lines of  sight through the  disk can transverse a  significant depth,
unless  the disk  is  infinitesimally thin.   Here  we illustrate  the
uncertainty  caused by a  $350\pc$ disk  scale-height on  the circular
velocity  predictions displayed  in Figure~7.   The contours  show the
levels where  the uncertainty  in the circular  velocity model  due to
this  disk  thickness attains  values  of  (from  the outside  in)  of
$10\kms$, $20\kms$  and $30\kms$. We include this  additional error in
when  calculating the  lag in  velocity of  stars behind  the circular
speed.   However,  over most  of  the  survey  region this  additional
uncertainty is less than $10\kms$.}
\end{figure}

\subsection{Inner fields}

We  now proceed  to  present the  velocity  data. The  upper panel  of
Figure~9 shows  the (Heliocentric)  radial velocity measurements  as a
function of $x$, the  major axis distance\footnote{We define the major
axis  distance  as  the  distance  from the  minor  axis  towards  the
North-east parallel to the major  axis.}  for the disk fields labeled
D1--D5  and which  lie within  the inner  ellipse of  Figures 1  and 4
(corresponding to  a disk  radius $R  < 27 \kpc$);  a small  number of
stars from field  F13 also fall into this  selection.  The dashed line
here  marks a  Heliocentric velocity  of $v  = -300\kms$,  the adopted
systemic velocity  of M31.  Non-rotating populations such  as might be
expected for  the stellar halo,  will scatter in a  broad distribution
around  this line.  In these  North-eastern fields,  the disk  lies at
radial  velocities  above  the  systemic  value,  so  counter-rotating
populations  can be identified  from their  location below  the dashed
line.  The bottom  panel shows the result of  subtracting the model of
Figure~7 from the  velocity data on a star by  star basis (rather than
using a single value for a  whole field).  Stars on circular orbits in
the plane of the disk  lie at a disk-corrected velocity of $v_{lag}=0$
(emphasized with a dashed line). Field D1 (at a major axis distance of
$x \sim  7\kpc$) appears to  be peculiar with an  intriguing kinematic
sub-structure at  Heliocentric velocity of $v \sim  -250\kms$; we will
discuss this structure in a subsequent contribution. It is interesting
to  note  that  this  sub-structure  was  only  detected  through  its
kinematic signature, as  there is no clear overdensity  in our maps of
RGB stars in this field.

However,  the disk  fields  D2--D5 (located  at  major axis  distances
between $12\kpc  < x  < 23\kpc$) show  a very similar  distribution of
stars, with a  broad halo component below $v_{lag}  = -150\kms$, and a
well-populated  peak  lagging   circular  orbits  only  slightly.   In
Figure~10 we show  the distribution of disk lag  velocities for fields
D2--D5, together with  a Gaussian fit of the  data between $-150\kms <
v_{lag} < 100\kms$, which has a mean of $\overline{v}_{lag} = -34\kms$
with dispersion  of $\sigma_v =  47\kms$ ($\overline{v^\prime}_{lag} =
-18\kms$, $\sigma_{v^\prime}=47\kms$).   The dash-dotted line  shows a
similar maximum-likelihood Gaussian fit, this time taking into account
the  estimated   velocity  uncertainty  on   the  individual  velocity
measurements,  and also  the uncertainty  in the  disk  rotation model
(from Figure~8).   This fit, which has  $\overline{v}_{lag} = -37\kms$
and  $\sigma_v  =   44\kms$  ($\overline{v^\prime}_{lag}  =  -23\kms$,
$\sigma_{v^\prime}=45\kms$),   shows  that   the  dispersion   of  the
population is  not artificially enhanced to any  significant extent by
our line of sight depth through these fields.

\begin{figure}
\begin{center}
\includegraphics[angle=270,width=\hsize]{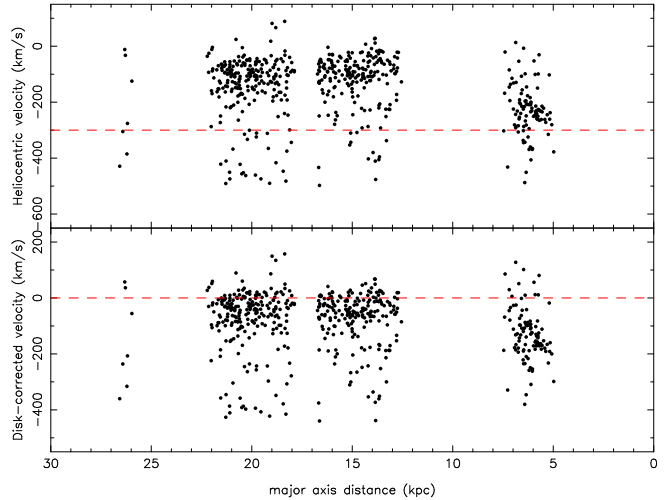}
\end{center}
\caption{Radial  velocity as  a function  of major  axis  distance for
fields D1--D5  (with a few  stars from F13  that lie within  the inner
ellipse of  Figure~4). The upper  panel shows the  Heliocentric radial
velocity  of the  stars, with  the  dashed line  marking the  systemic
velocity of  M31. The bottom  panel shows the same  data ``corrected''
for the  expected disk rotation given  by the model  shown in Figure~7;
now the  dashed line marks the  location of stars  on circular orbits.
Note the obvious structure that is  present in the field at major axis
distance of $\sim 7\kpc$ (field D1).}
\end{figure}

\begin{figure}
\begin{center}
\includegraphics[angle=270,width=\hsize]{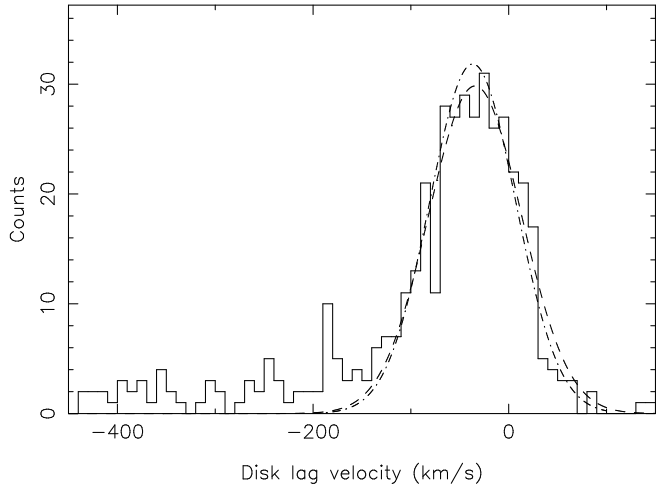}
\end{center}
\caption{The  summed distribution  of velocity  lags of  fields D2--D5
(i.e. stars with major axis  distances between $12\kpc$ to $23\kpc$ in
Figure~9). The large  peak between $-150\kms < v_{lag}  < 100\kms$ can
be fit  with a  Gaussian (dashed line)  of mean  $\overline{v}_{lag} =
-34\kms$ and  dispersion $\sigma_v=47\kms$ ($\overline{v^\prime}_{lag}
=  -18\kms$,  $\sigma_{v^\prime}=47\kms$).  The  fit  does not  change
significantly if  we take into  account the estimated  velocity errors
and the  disk rotation model  uncertainties displayed in  Figure~8, we
obtain  then  (dot-dashed   line)  $\overline{v}_{lag}  =  -37  \kms$,
$\sigma_v=44\kms$      ($\overline{v^\prime}_{lag}     =     -23\kms$,
$\sigma_{v^\prime}=45\kms$).}
\end{figure}

As  discussed  above,  contamination  from  Milky  Way  stars  can  be
problematic for  $v >  -100\kms$; however the  M31 disk has  very high
contrast in these inner fields (D1--D5), with a source surface density
in our CMD  selection window that is 20 times higher  than in the halo
field F7, so we expect the  Galactic contamination to be small, 5\% at
the very most.  Thus we find that the dominant population in the inner
regions  of  M31  is a  thin  disk  population  with a  dispersion  of
$\sigma_v = 44\kms$ that trails circular orbits by $\overline{v}_{lag}
= -37\kms$; these values are reasonably similar to the measurements of
the   velocity  dispersion  of   the  Galactic   disk  in   the  Solar
Neighborhood    where    the     velocity    ellipsoid    has    axes
$(\sigma_R,\sigma_\phi,\sigma_z) =  (38:24:17)\kms$ and the population
lags the circular  velocity with an asymmetric drift  of $\sim 20\kms$
\citep{dehnen98}.  The asymmetric drift  of the M31 population will be
discussed further  in \S6.1.  Note that the  very inclined orientation
of M31  means that we  do not have  access to the  vertical dispersion
$\sigma_z$. The  measured $\sigma_v$  is a mixture  of the  radial and
azimuthal  components $\sigma_R$ and  $\sigma_\phi$, dependent  on the
field under study.  For simplicity  we neglect this distinction in the
present discussion, referring to a single $\sigma_v$.

\begin{figure}
\begin{center}
\includegraphics[angle=270,width=\hsize]{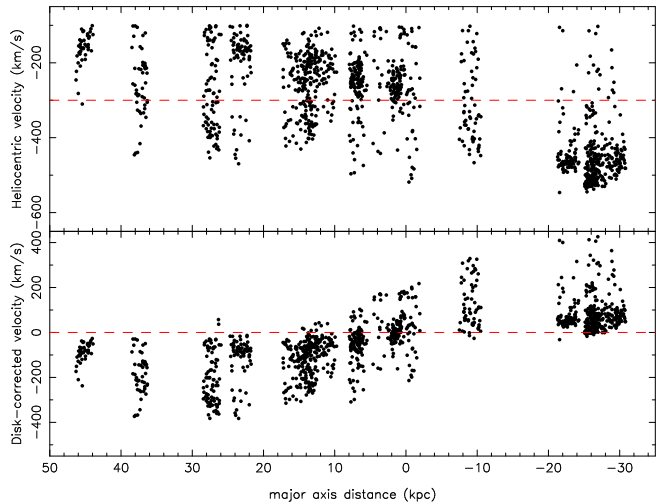}
\end{center}
\caption{As Figure~9, but  for the 16 fields F1--F16  that lie outside
of the inner ellipse shown in Figures 1 and 4, and which no not lie in
the giant stream.  In most of these fields  the dominant population of
stars  is  seen  with  velocities  lagging the  circular  velocity  by
velocities between $0\kms$ to $\sim 100\kms$.}
\end{figure}

\subsection{Outer fields}

Having  shown that  the  inner  fields have  kinematics  typical of  a
canonical thin disk, we now  examine the outer field data.  The fields
lying on the giant stellar stream (those prefixed by the letter 'S' in
Figure~4) are removed from the sample for the sake of simplicity. This
leaves a total of 16 fields marked F1--F16 on Figure~4.

The  measured radial  velocities in  these fields  are displayed  as a
function of  major axis  distance on the  top panel of  Figure~11. For
clarity, we have removed from  this sample all stars with heliocentric
velocities   $v   >  -100\kms$,   thereby   minimizing  the   Galactic
contamination (\cf\ Galactic model in  Figure~5). There is a wealth of
kinematic substructure in these fields (Chapman \etal\ 2005, in prep),
and a  significant halo population seen scattered  around the systemic
velocity  (dashed  line); however  the  dominant  population in  these
fields again follows  closely the disk rotation.  This  can be seen on
the bottom panel of Figure~11, where we have corrected the data on the
upper  panel for the  rotation model  of Figure~7.   In the  fields at
major  axis distance  between $-20\kpc  < x  < -30\kpc$,  the numerous
population stands out clearly with ``disk-corrected velocity'' between
$0\kms$  and  $100\kms$. On  this  South-western  side  of M31,  these
velocities  correspond to  a lagging  behind the  \ion{H}{1}.   On the
opposite side  of the  disk, at  major axis distances  $x >  5\kpc$, a
similar  concentration  is   found  with  ``disk-corrected  velocity''
between $-100\kms$  and $0\kms$, this  again corresponds to  a lagging
velocity behind the \ion{H}{1} gas.

For completeness,  in Figure~12  we show the  same velocity data  as a
function  of minor  axis  distance\footnote{We define  the minor  axis
distance as  the distance from the  major axis towards  the North-west 
parallel to the minor axis.}, $y$.  It is clear from this diagram that
the halo fields F5 and F7 (the  two fields at minor axis distance $y <
-15\kpc$) do  not contribute significantly  (if at all) to  the peaked
velocity structure that rotates as a disk.

\begin{figure}
\begin{center}
\includegraphics[angle=270,width=\hsize]{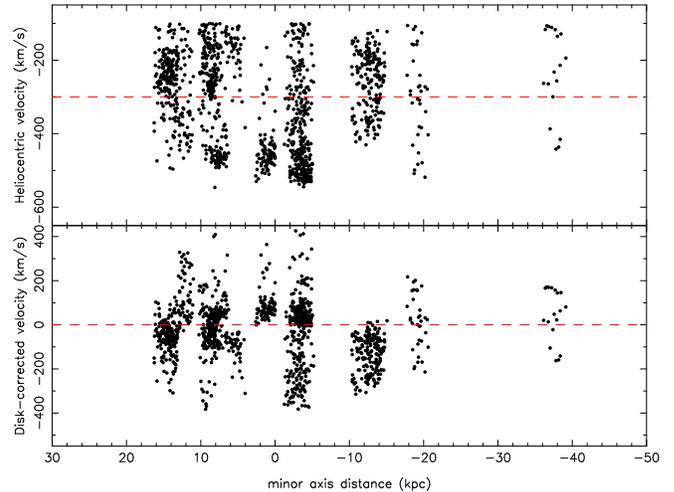}
\end{center}
\caption{As Figure~11, but showing the velocity as a function of minor
axis distance.}
\end{figure}

Figure~13 shows the  distribution of velocity lag $v_{lag}$  in the 16
fields under consideration. The velocity lag is defined to be equal to
the  disk-corrected  velocity  for   fields  at  positive  major  axis
distance, but is inverted for  fields at negative major axis distance.
Each  panel is  labeled with  the  field name  and as  well as  three
numbers  which  give the  major  axis  distance  $x$, the  minor  axis
distance $y$ and the de-projected  M31-plane radius $R$, of the centre
of each field,  all in kpc. A striking  feature of these distributions
is the  presence of a  narrow component in  most of the  fields, which
lags the circular  velocity by merely $\sim 50\kms$, that  is to say a
population which  has distinctly disk-like kinematics.  Yet  this is a
startling finding, since these  fields are extremely distant from M31,
lying   at   radial  distances   between   $30\kpc$   and  more   than
$50\kpc$. That this behavior is the most salient kinematic feature of
the survey is demonstrated  in Figure~14, which shows the distribution
of velocity lag as a function of the M31-plane radial distance, R, for
all stars in  the survey in non-stream fields with  $R < 80\kpc$.  The
majority of  the stars are found  with disk lag velocity  in the range
$-100\kms < v_{lag} < 0\kms$, and therefore take part in a large-scale
rotational motion.

\begin{figure}
\begin{center}
\includegraphics[angle=0,width=\hsize]{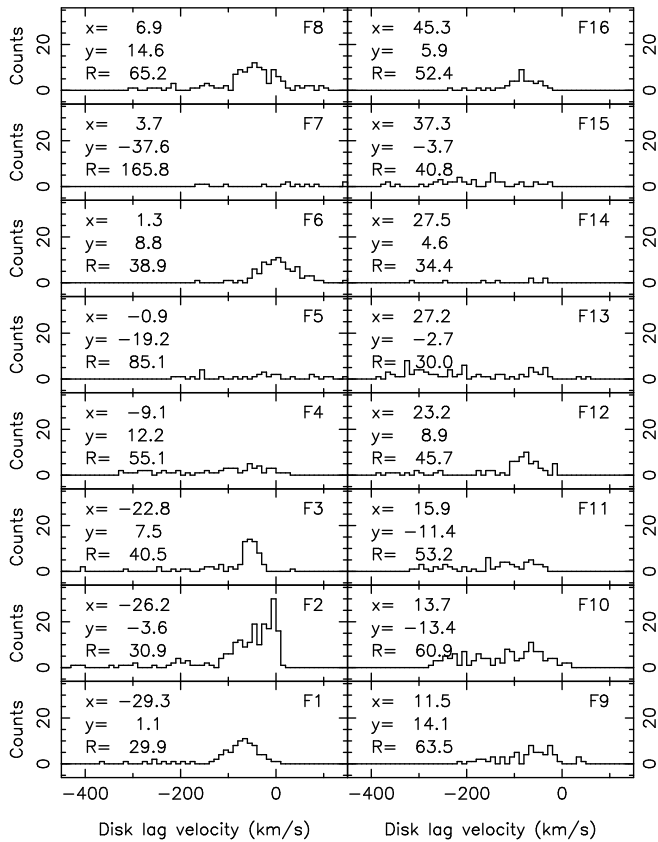}
\end{center}
\caption{Distributions  of  the   velocity  lag  behind  the  circular
velocity model for the 16 fields displayed in Figure~11. In each panel
the three  numbers show, respectively,  from top to bottom,  the major
axis distance,  the minor axis  distance, and the  de-projected radius
$R$ of the fields.  The field name is also indicated.}
\end{figure}

\begin{figure}
\begin{center}
\includegraphics[angle=270,width=\hsize]{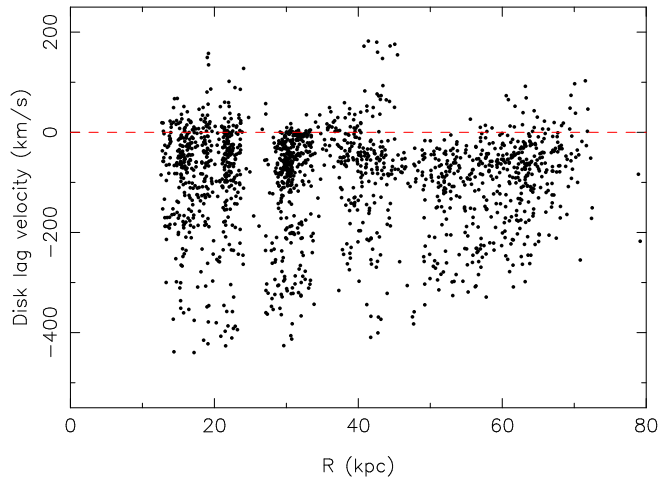}
\end{center}
\caption{The distribution of velocity  lags as a function of M31-plane
de-projected radial distance  in all non-stream survey fields  at $R <
80\kpc$.  The majority  of  these stars  display  small velocity  lags
($-100\kms <  v_{lag} < 0\kms$),  and are therefore  rotating rapidly.
The dashed line shows the location of circularly-rotating stars.}
\end{figure}

Summing  the velocity lag  distributions in  all the  outer non-stream
fields (except  for F6 to  remove any concern  regarding contamination
from NGC~205  --- this will be  discussed further in  \S5.4) gives the
distribution shown in Figure~15.  This distribution is very similar to
that  previously constructed  for the  inner fields  (\cf\ Figure~10),
showing again  a broad halo distribution  with the addition  of a very
strong peak,  in this case  centered at $\overline{v}_{lag}  = -54\kms$
with a dispersion of $\sigma_v = 37\kms$ ($\overline{v^\prime}_{lag} =
-41\kms$, $\sigma_{v^\prime} = 38\kms$).

\begin{figure}
\begin{center}
\includegraphics[angle=270,width=\hsize]{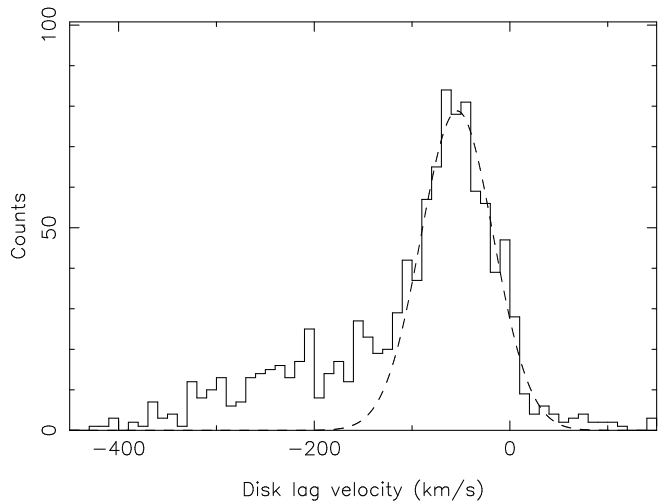}
\end{center}
\caption{The  summed distribution of  velocity lags  in all  fields in
Figure~13, except F6.  A Gaussian  fit to the data between $-150\kms <
v_{lag} <  100\kms$ is superimposed  (dashed line), which has  mean at
$\overline{v}_{lag} = -54\kms$ and a dispersion of $\sigma_v = 37\kms$
($\overline{v^\prime}_{lag}    =   -41\kms$,    $\sigma_{v^\prime}   =
38\kms$).}
\end{figure}

These data unambiguously show that the bulk of the stars in our survey
fields  share a  common kinematic  property: they  co-rotate  with the
\ion{H}{1}  gas, with velocities  that lag  that of  the gas  by $\sim
40\kms$, and  possess a velocity  dispersion of $\sim  40\kms$.  There
therefore appears to be a  vast stellar disk-like structure around the
Andromeda galaxy, distributed on an unprecedented scale.

\section{Overview of selected outer fields}

We  next present  in  detail the  data  from 7  of  the outer  fields,
selected  from Figure~13  for having  a particularly  strong disk-like
peak in their velocity distributions.  The selected fields are F1, F2,
F3, F6, F8, F12 and F16.

\subsection{Field F1}

Field F1  lies at the  South-western boundary of the  conspicuous dark
and messy structure  in Figure~1 that surrounds the  bright inner disk
seen  in photographic  surveys.  The  most  massive of  the M31  outer
globular clusters, named ``G1'' also  lies in this field, and we refer
to the structure  at this location as the G1  clump.  We summarize the
structural,  kinematic   and  photometric   data  on  this   field  in
Figure~16. The upper  left-hand panel gives the location  of the field
with respect to the large survey area, which we zoom into in the upper
right  hand   panel.   The  bottom  left  panel   shows  the  velocity
distribution of the stars in our RGB selection window, together with a
Gaussian fit centered at $\overline{v}_{lag} = -69\kms$ with dispersion
$\sigma_v   =    30\kms$   ($\overline{v^\prime}_{lag}   =   -51\kms$,
$\sigma_{v^\prime} =  29\kms$). The bottom right-hand  panel shows the
CMD derived from the INT survey over the $16.7\arcmin \times 5\arcmin$
field of  view of the  DEIMOS instrument. Those stars  with velocities
within $2\sigma$  of the mean  of the Gaussian  fit are marked  with a
filled circle.  The four curves  superimposed on this CMD are globular
cluster  fiducials, with metallicities,  from left  to right  of ${\rm
[Fe/H] =  -1.91 }$, ${\rm [Fe/H] =  -1.29 }$, ${\rm [Fe/H]  = -0.71 }$
and ${\rm [Fe/H] = -0.2 }$.  The stars that partake in the strong peak
evidently  scatter in  a wide  distribution around  ${\rm  [Fe/H] \sim
-0.7}$.  Both   the  kinematics  and   the  spectroscopic  metallicity
(presented in  \S5.9) that  we derive in  this field are  in excellent
agreement with the findings of \citet{reitzel04}.

\begin{figure}
\begin{center}
\includegraphics[bb= 53 57 540 660, clip, angle=0,width=\hsize]{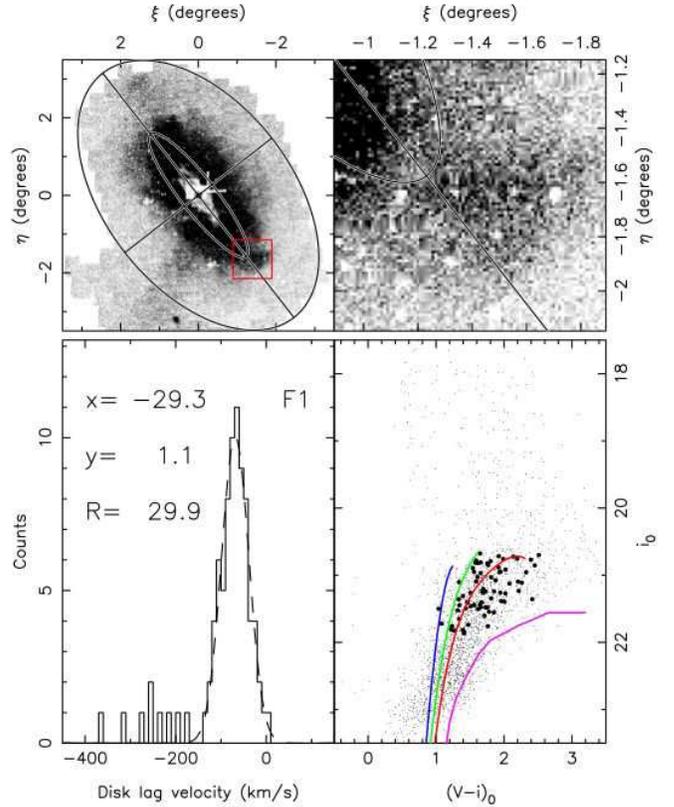}
\end{center}
\caption{The upper  left panel shows the  location of the  field F1 in
our survey region  (compare to Figure~1), with a  $1\deg \times 1\deg$
box  marking the  location  of  the zoomed-in  view  displayed on  the
right-hand  panel, which  is  centered  on the  centre  of this  DEIMOS
field. The  white holes present  in this map  are due to the  halos of
bright stars.   The bottom left-hand  panel shows the  distribution of
the velocity lag behind the  circular velocity in this field, together
with   a  single-component  Gaussian   fit,  which   has  a   mean  of
$\overline{v}_{lag} =  -69\kms$ and dispersion of  $\sigma_v = 30\kms$
($\overline{v^\prime}_{lag} = -51\kms$, $\sigma_{v^\prime} = 29\kms$).
The  bottom-right  panel shows  the  color-magnitude  diagram of  the
field, with small dots showing the full INT photometry within the area
of the  DEIMOS field of  view. The filled  circles are the  stars with
observed velocities within $\pm 2\sigma$ of the fitted Gaussian on the
bottom  left-panel.   For comparison,  we  have superimposed  fiducial
cluster sequences for (from  left to right) NGC~6397, NGC~1851, 47~Tuc
and NGC~6553,  which have  metallicities of ${\rm  [Fe/H] =  -1.91 }$,
${\rm [Fe/H]  = -1.29 }$, ${\rm [Fe/H]  = -0.71 }$ and  ${\rm [Fe/H] =
-0.2 }$, respectively.}
\end{figure}

The spatial structure  of this grouping of stars  is very interesting.
An inspection  of the top  right-hand panel (and also  Figure~1) shows
that this structure is an irregular lump -- somewhat elongated E-W and
apparently distinct from the disk.  The structure is very large, being
more than $10\kpc$ long. It  is highly unlikely that this structure is
bound, as this  would require some $\sim 10^{9}\msun$  of dark matter,
given  the velocity  dispersion and  extent.  Instead,  the disorderly
morphology argues in favor of this being tidal debris.

\subsection{Field F2}

Field F2 (Figure~17) is located close to Field F1, but in a prominence
that descends  southwards of the  disk.  The velocity  distribution in
field F2  is similar to that  of Field F1, but  contains an additional
narrow component, which  is seen as a spike near $v_{lag}  = 0$ in the
bottom left  hand panel of Figure~17.  Fitting  two Gaussian functions
yields a narrow peak  at $\overline{v}_{lag} = -4\kms$ with dispersion
$\sigma_v    =   7\kms$    ($\overline{v^\prime}_{lag}    =   11\kms$,
$\sigma_{v^\prime}    =   5\kms$),   and    a   broader    peak   with
$\overline{v}_{lag}    =   -46\kms$    and    $\sigma_v   =    35\kms$
($\overline{v^\prime}_{lag} = -29\kms$, $\sigma_{v^\prime} = 33\kms$).
The stars that gives rise  to the narrow kinematic spike are dispersed
spatially over the whole DEIMOS  field, so the structure is not simply
a cluster.  The small velocity dispersion and the mean velocity almost
exactly  on  the  expected   circular  rotation  imply  that  the  the
population  is almost certainly  young.  The  population in  the broad
kinematic peak  is again very similar  to the peak that  we observe in
field F1 and in the inner fields D2--D5.

\begin{figure}
\begin{center}
\includegraphics[bb= 53 57 540 660, clip, angle=0,width=\hsize]{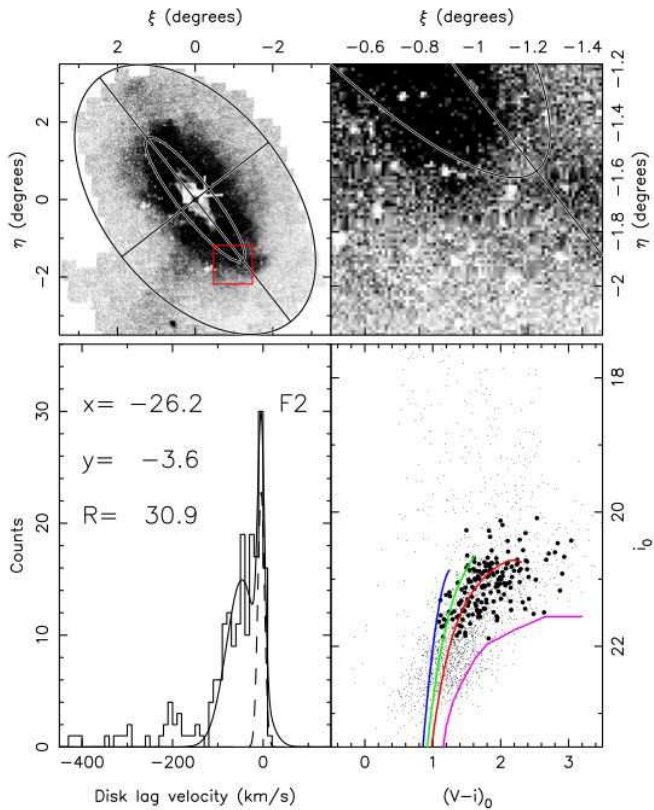}
\end{center}
\caption{As Figure~16, but  for field F2.  In this case  we have fit a
two-component  Gaussian  model:  we  find  a narrow  peak  centered  at
$\overline{v}_{lag} =  -4\kms$ with  dispersion of $\sigma_v  = 7\kms$
($\overline{v^\prime}_{lag}  = 11\kms$, $\sigma_{v^\prime}  = 5\kms$),
and a broader peak with $\overline{v}_{lag} = -46\kms$ and $\sigma_v =
35\kms$  ($\overline{v^\prime}_{lag} = -29\kms$,  $\sigma_{v^\prime} =
33\kms$).}
\end{figure}

\subsection{Field F3}

Field  F3  is  located towards  the  South-west  of  M31, on  a  dense
elongated  structure that  trails behind  the disk  in a  way  that is
reminiscent  of  a tidal  stream  or  the end  of  a  spiral arm  (see
Figure~18).  Though the velocity  dispersion of $\sigma_v = 15\kms$ is
relatively  low in this  field, the  color-magnitude position  of the
stars that populate the kinematic peak is very similar to that seen in
other fields.

\begin{figure}
\begin{center}
\includegraphics[bb= 53 57 540 660, clip, angle=0,width=\hsize]{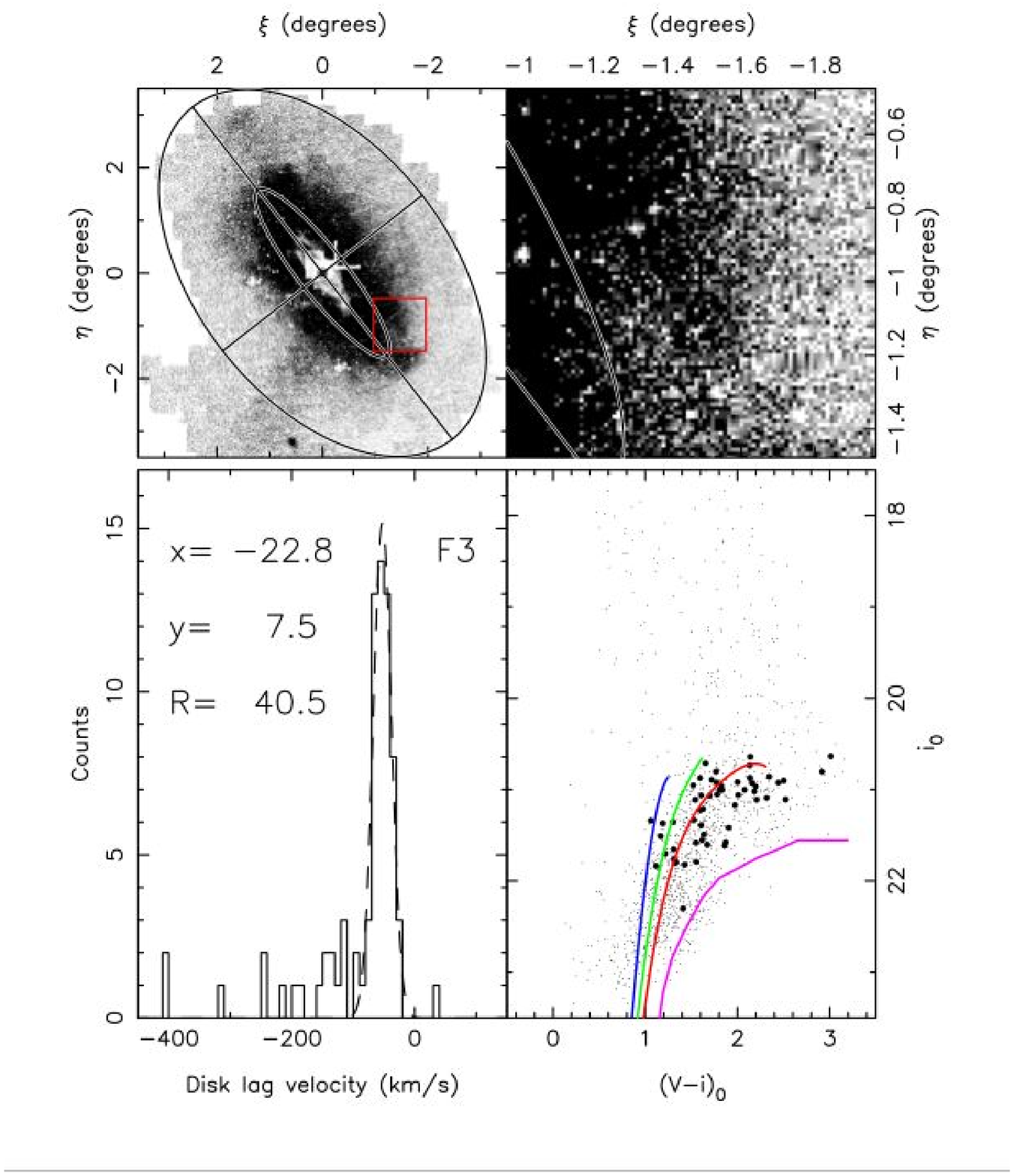}
\end{center}
\caption{As  Figure~16, but  for  field F3.   The  Gaussian model  has
$\overline{v}_{lag}     =-53\kms$    and    $\sigma_v     =    15\kms$
($\overline{v^\prime}_{lag}    =   -37\kms$,    $\sigma_{v^\prime}   =
15\kms$).}
\end{figure}

\subsection{Field F6}

Figure~19 shows  the velocity and color-magnitude data  for field F6,
centered on the  Northern edge of NGC~205, one  of the dwarf elliptical
satellite galaxies of Andromeda.  The distribution of velocities has a
narrow peak  centered at  $\overline{v}_{lag} = 2\kms$  with dispersion
$\sigma_v    =   36\kms$    ($\overline{v^\prime}_{lag}    =   4\kms$,
$\sigma_{v^\prime} = 37\kms$). Surprisingly, this distribution is very
similar  to  what has  been  found  previously  in other  fields:  the
velocity dispersion is small and consistent with a circularly rotating
population.  Figure~20 displays the  velocity and spatial data in more
detail,  with  the middle  panel  showing  the  target positions  with
respect to a sketch of NGC~205. The upper panel gives the heliocentric
velocity measurements,  while the bottom panel shows  the velocity lag
as a function of standard coordinate $\xi$.  It is interesting to note
that if  some stars in this  field were being removed  from NGC~205 by
the  action  of tidal  forces,  they  would  contribute, at  least  in
projection,  to stars  with disk-like  kinematics; we  return  to this
point in \S6\ below.

\begin{figure}
\begin{center}
\includegraphics[bb= 53 57 540 660, clip, angle=0,width=\hsize]{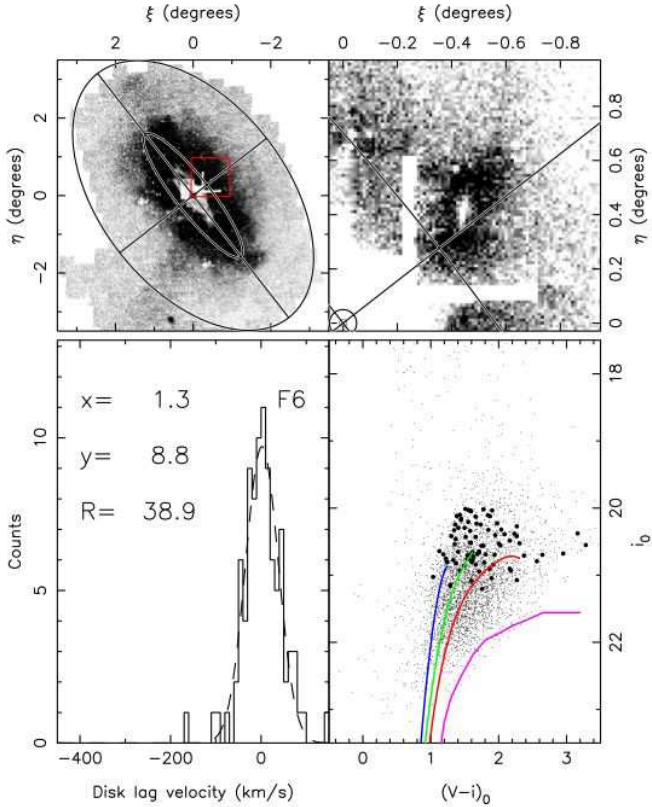}
\end{center}
\caption{As  Figure~16, but  for  field F6.   The  Gaussian model  has
$\overline{v}_{lag}    =    2\kms$    and    $\sigma_v    =    36\kms$
($\overline{v^\prime}_{lag}  = 4\kms$, $\sigma_{v^\prime}  = 37\kms$).
The  white  ``L''-shaped  hole  in  the  finding chart  is  due  to  a
deliberate shifting of  the camera position to centre  one of the CCDs
of the mosaic on NGC~205; part  of the inner survey region was thereby
not covered.  The hole in the  centre of NGC~~205 is  a consequence of
crowding.}
\end{figure}

\begin{figure}
\begin{center}
\includegraphics[angle=0,width=\hsize]{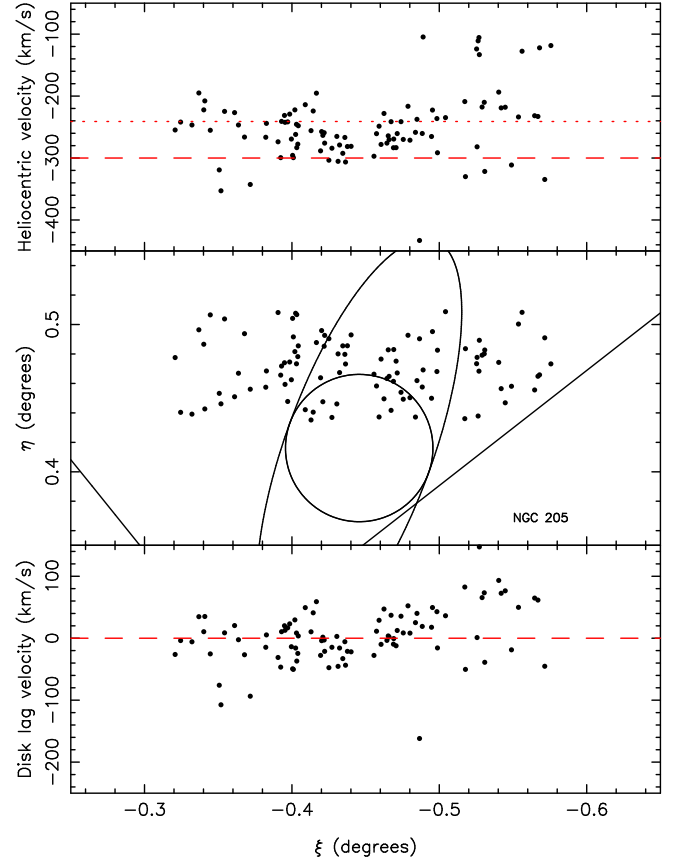}
\end{center}
\caption{The central panel shows the positions of the targets in field
F6 with  respect to  a sketch of  NGC~205. The diagonal  line segments
show the minor  and major axes of M31. The top  and bottom panels show
the kinematic data as a function of the standard coordinate $\xi$. The
dashed line in the top panel marks the systemic velocity of M31, while
the dotted  line shows the  systemic velocity of NGC~205.   The bottom
panel displays  the corresponding  distribution of disk  lag velocity,
with the dashed line marking the locus of circular orbits.}
\end{figure}

\subsection{Field F8}

Figure~21 shows the  data for field F8, a field  in the possible tidal
trail  of  NGC~205.   This   structure  was  discussed  previously  by
\citet{mcconnachie04b},  where they  presented the  morphology  of the
feature, together with kinematics from our field F9 (their field W91).
The  new data  in field  F8 (adjacent  to field  F9) do  not  show the
bimodal distribution  found by \citet{mcconnachie04b},  which suffered
from  a  small sample  size.   Instead  the  only significant  stellar
population  present  has,  yet  again,  disk-like  kinematics  with  a
dispersion  of $\sigma_v  = 38\kms$  centered at  $\overline{v}_{lag} =
-43\kms$ ($\overline{v^\prime}_{lag}  = -37\kms$, $\sigma_{v^\prime} =
35\kms$).

\begin{figure}
\begin{center}
\includegraphics[bb= 53 57 540 660, clip, angle=0,width=\hsize]{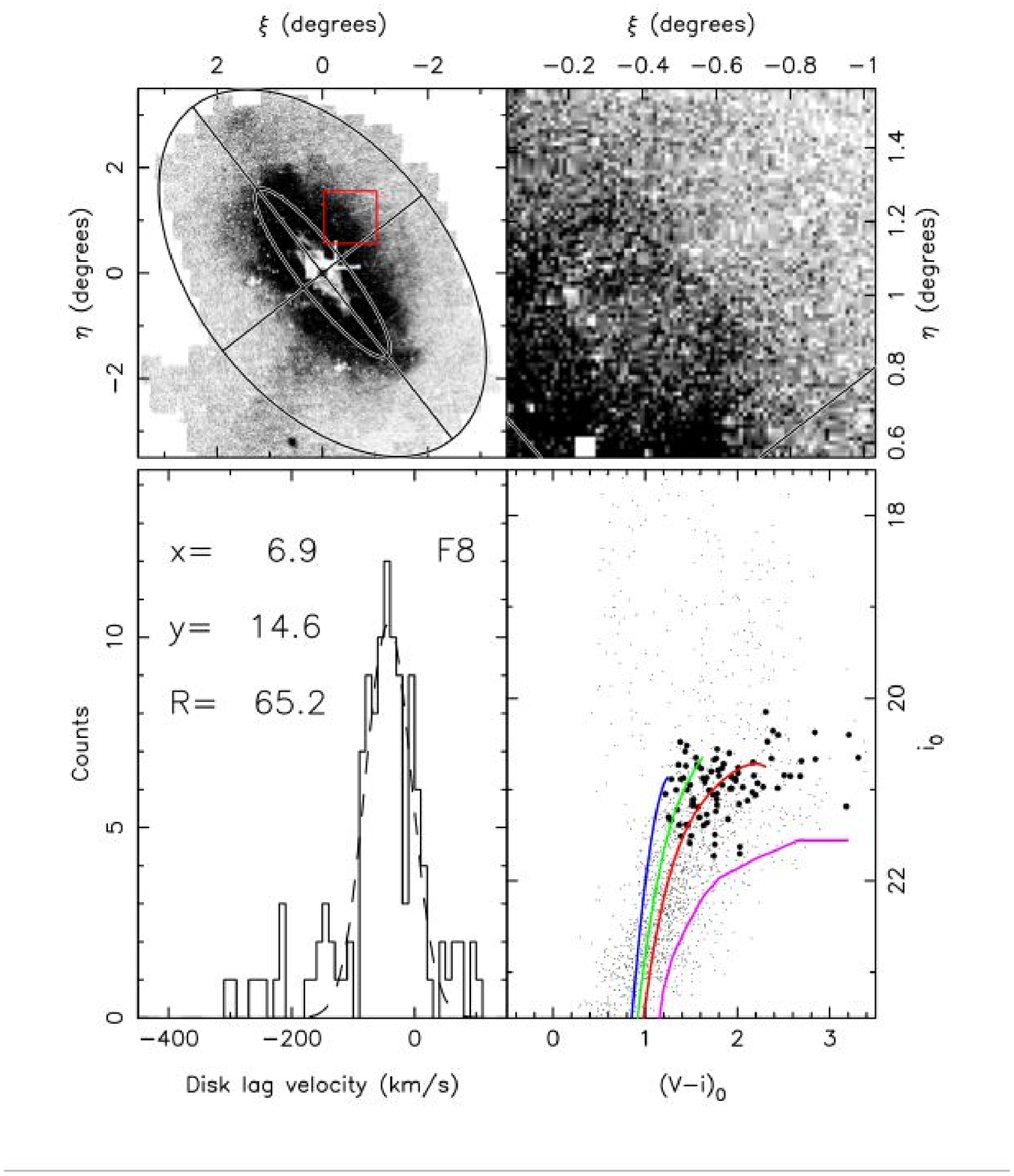}
\end{center}
\caption{As  Figure~16,  but for  field  F8.  The  Gaussian model  has
$\overline{v}_{lag}    =   -43\kms$    and    $\sigma_v   =    38\kms$
($\overline{v^\prime}_{lag}    =   -37\kms$,    $\sigma_{v^\prime}   =
35\kms$).}
\end{figure}

\subsection{Field F12}

Figure~22  shows the location  of another  dense structure,  this time
located  towards   the  North-eastern  end  of  the   major  axis.   A
peculiarity of  this feature (the ``Northern Spur'')  is its curious
wedge-shaped morphology  (seen in the upper  right-hand diagram) which
displays  a  very sharp  boundary.   The  velocity  dispersion of  the
structure $\sigma_v = 24\kms$  ($\sigma_{v^\prime} = 23\kms$) is again
similar to that in other fields.

\begin{figure}
\begin{center}
\includegraphics[bb= 53 57 540 660, clip, angle=0,width=\hsize]{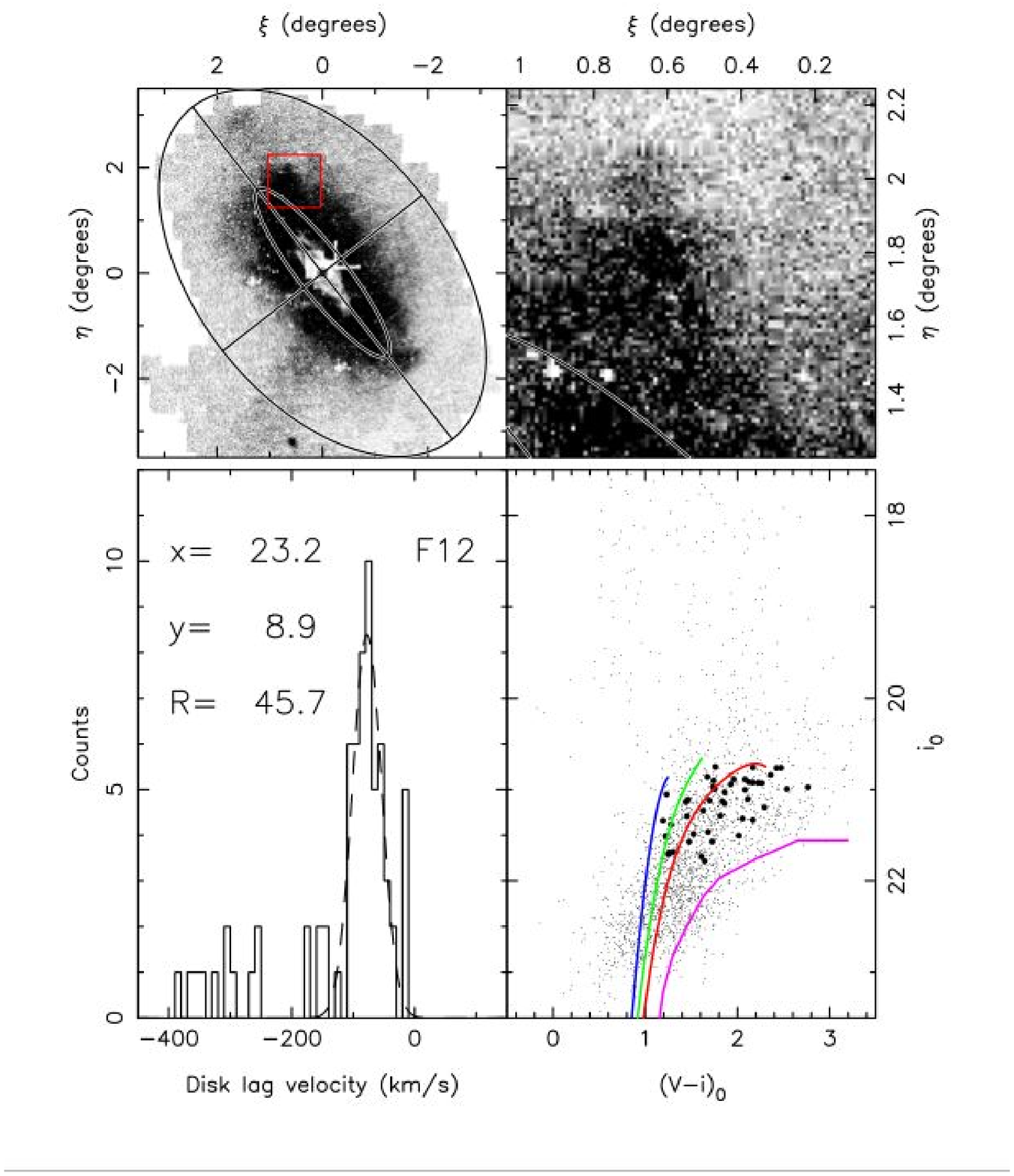}
\end{center}
\caption{As  Figure~16, but  for field  F12.  The  Gaussian  model has
$\overline{v}_{lag}    =   -78\kms$    and    $\sigma_v   =    24\kms$
($\overline{v^\prime}_{lag}    =   -60\kms$,    $\sigma_{v^\prime}   =
23\kms$).}
\end{figure}

\subsection{Field F16}

Field F16 is  situated in a particularly interesting  location, on the
North-eastern end of our survey  region (see Figure~23), where our INT
survey detected a very extended very low surface brightness structure.
It  lies  at the  greatest  projected  radius  ($46\kpc$) of  all  the
non-stream fields  we observed.  \cite{zucker} argue  that this object
(which they call Andromeda NE) may be one of the most low mass and low
surface  brightness  galaxies  found  to  date  or  possibly  torn-off
material from the disk of M31. The radial velocity distribution in the
bottom left-hand  panel of Figure~23 shows a  well-defined narrow peak
of  dispersion $\sigma_v  = 27\kms$  centered at  $\overline{v}_{lag} =
-80\kms$ ($\overline{v^\prime}_{lag}  = -65\kms$, $\sigma_{v^\prime} =
28\kms$).  However, this field is the  only one in our survey in which
there  could be  some concern  about Galactic  contamination affecting
significantly  the fitted disk-like  peak.  In  Figure~24 we  show the
Heliocentric velocity  distribution of stars  in this field,  with the
cut at $v = -100\kms$, imposed  on all fields to minimize the Galactic
contaminants.  The  predictions of the Besan{\c c}on  Galaxy model are
superposed  (green histogram)  normalised to  the  $16.7\arcmin \times
5\arcmin$  field of view  of DEIMOS,  but corrected  for the  ratio of
available  to observed  targets  in our  CMD  selection window.   Thus
according  to   this  model,  below   $v  =  -100\kms$   the  Galactic
contribution  will  be  minor,  and  can  be  neglected.   The  bottom
right-hand  panel of  Figure~23 shows  the CMD  location of  the stars
within the  peak; evidently these are relatively  metal-rich, and very
similar  to the  other disk-like  stars found  in the  fields examined
previously.

The  velocity histogram  in the  bottom left-hand  panel  of Figure~23
shows a spike  at velocities $-90\kms < v_{lag}  < -80\kms$. The stars
partaking in this spike are  uniformly distributed over the field, and
they have  a similar color  spread to the  other stars in  the sample
(the  corresponding  points are  circled  in  the  CMD on  the  bottom
right-hand panel).   This lack  of distinct properties,  together with
the  low statistics  indicate  that  this spike  is  most probably  an
insignificant ($\sim 2\sigma$) random deviation.

\begin{figure}
\begin{center}
\includegraphics[bb= 53 57 540 660, clip, angle=0,width=\hsize]{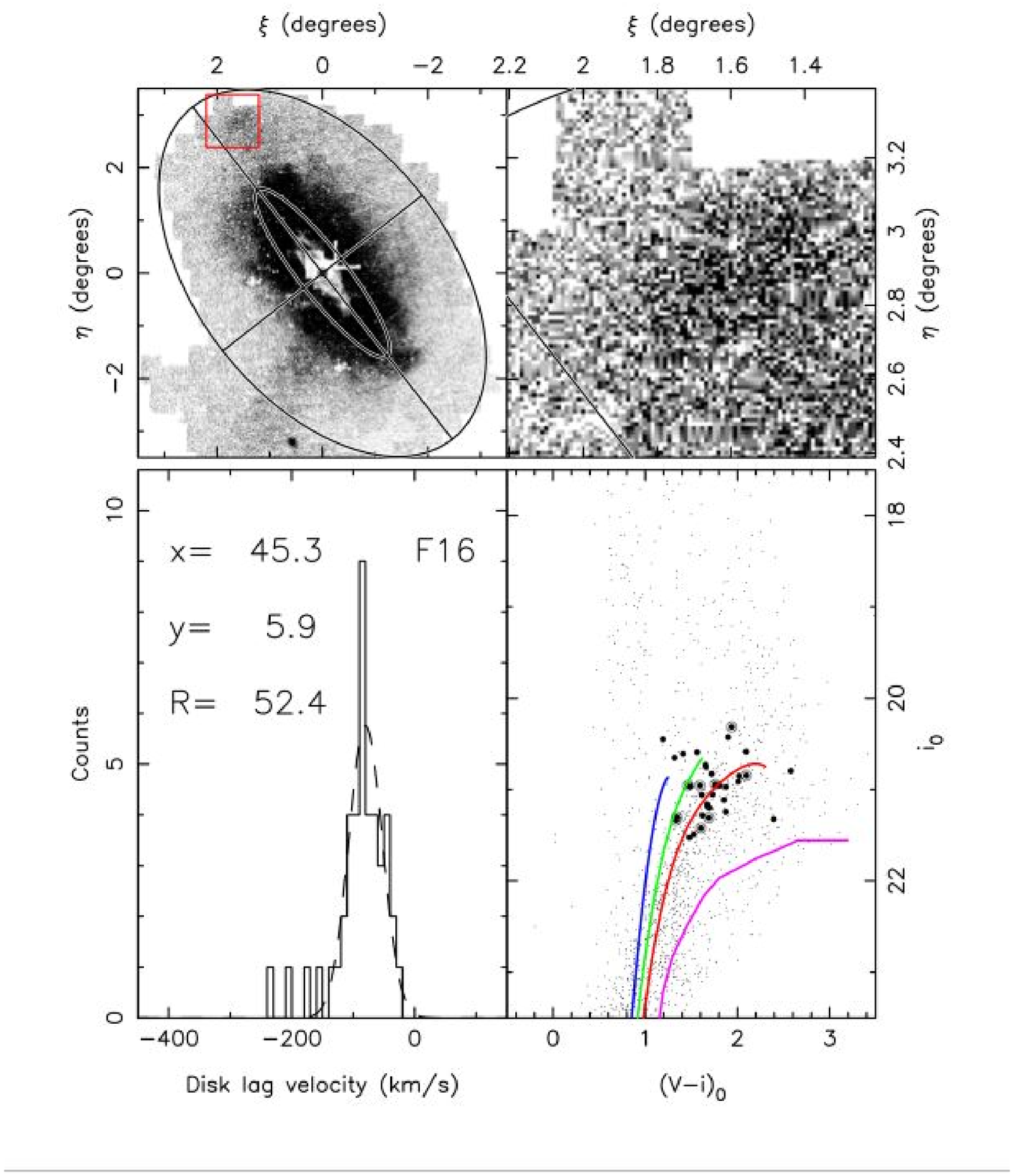}
\end{center}
\caption{As  Figure~16, but  for field  F16.  The  Gaussian  model has
$\overline{v}_{lag}    =   -80\kms$    and    $\sigma_v   =    27\kms$
($\overline{v^\prime}_{lag}    =   -65\kms$,    $\sigma_{v^\prime}   =
28\kms$).}
\end{figure}

\begin{figure}
\begin{center}
\includegraphics[angle=270,width=\hsize]{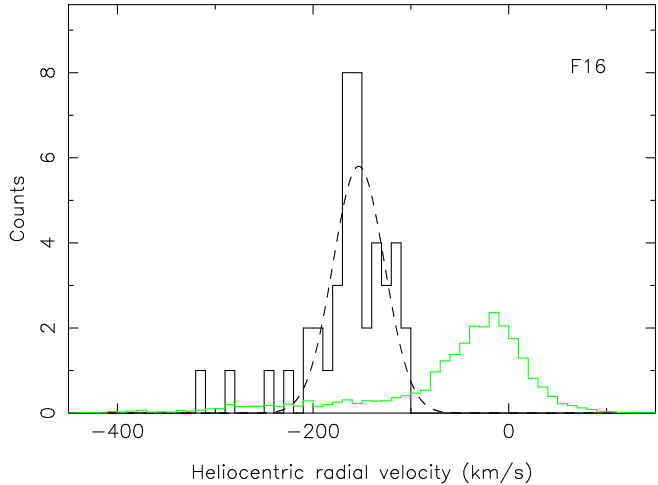}
\end{center}
\caption{The heliocentric velocity distribution of field F16, situated
in the  low-surface brightness overdensity  to the North-east  of M31.
The green histogram shows the prediction of the Besan{\c c}on model in
an area of  the DEIMOS field of view, but corrected  for the fact that
only 92  stars in  the CMD  selection window were  observed (out  of a
possible  282). The Besan{\c  c}on model  therefore predicts  that the
Galactic  contamination  is  minimal  at heliocentric  velocities  $v<
-100\kms$, which strengthens the results presented in Figure~23.}
\end{figure}

\subsection{Fields F4, F5, F7, F9, F10, F11, F13, F14 and F15}

The remaining outer non-stream fields:  F4, F5, F7, F9, F10, F11, F13,
F14 and F15, do not  individually possess a narrow and strong velocity
peak near  the expected velocity  of a circularly  rotating population
(see  Figure~13).  However,  by  summing the  fields  together we  can
enhance the signal  to detect a global population  which lags the disk
by a similar velocity. The  nine fields that were added together cover
a very  large area  over our survey.   Some fields, in  particular the
halo field F7,  are almost certainly devoid of any  disk stars, so the
``disk  velocity  lag''  is   not  a  meaningful  quantity,  and  will
contribute  only  noise  to  the summed  distribution.   Nevertheless,
Figure~25 shows  that a narrow peak  is again detected,  albeit with a
large population  of ``contaminants'', this time  stemming mostly from
the halo population  of M31 itself. Due to  the kinematic substructure
in  the halo  population (which  will  be presented  in a  forthcoming
paper, Chapman  et al.,  in prep.)  we  cannot easily remove  the halo
``contamination''  to  enhance  the  signal  of  the  disk-like  peak.
Nevertheless, to help guide the eye, we have used a maximum likelihood
algorithm to  fit a  Gaussian function to  the velocity  data windowed
between $-80\kms <  v_{lag} < 50\kms$, which we find  to be centered on
$\overline{v}  =  -55\kms$  with   dispersion  $\sigma_v  =  33  \kms$
($\overline{v^\prime}_{lag}    =   -41\kms$,    $\sigma_{v^\prime}   =
31\kms$). Although the reliability  of this fit is clearly compromised
by the  strong halo  contamination, the similarity  of this  peak with
those presented in  Figures~16 to 23 is very  striking.  This confirms
that  a disk-like  population is  present in  almost all  the surveyed
fields.

\begin{figure}
\begin{center}
\includegraphics[angle=270,width=\hsize]{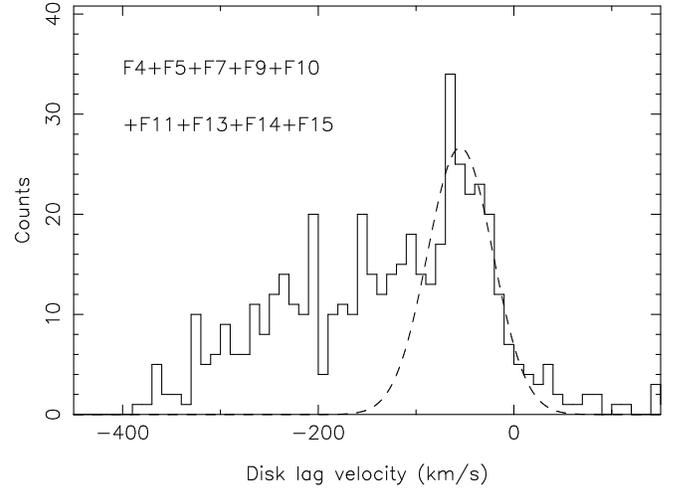}
\end{center}
\caption{The distribution of disk  lag velocities for the (non-stream)
external  fields   without  a  prominent   narrow  disk-like  velocity
component; these are: F4, F5, F7, F9, F10, F11, F13, F14 and F15. This
combining brings out  of the noise, yet again,  a narrow velocity peak
close to  expected velocity of  a circularly rotating  population. For
illustrative  purposes  only, we  have  fitted  the velocity  interval
$-80\kms < v_{lag}  < 50\kms$ with a Gaussian  of mean $\overline{v} =
-55\kms$     and      dispersion     $\sigma_v     =      33     \kms$
($\overline{v^\prime}_{lag}    =   -41\kms$,    $\sigma_{v^\prime}   =
31\kms$).}
\end{figure}

\subsection{Spectroscopic metallicities}

The  spectra  also allow  a  measurement  of  the metallicity  of  the
targeted stars  from the equivalent  widths (EW) of the  Ca~II triplet
absorption lines.  While the  noise in individual spectra is typically
too high to yield a useful measurement of the Ca~II equivalent widths,
we proceeded to  estimate the average metallicity by  stacking the RGB
star spectra  of the targets within  $2\sigma$ of the  fitted peaks of
the fields displayed in Figures  16 to 23 (the corresponding stars are
also highlighted in the CMD  panels).  The spectra are shifted to zero
velocity before stacking.  We follow as closely as possible the method
of \citet{rutledge}, fitting Moffat  functions to the Ca~II lines. The
average spectrum yields a  measurement of the Ca~II triplet equivalent
width  used to  estimate the  metallicity as  $[Fe/H] =  -2.66  + 0.42
[\Sigma  Ca  -  0.64 (V_{HB}  -  V_{ave})]$,  with  $\Sigma Ca  =  0.5
EW_{\lambda 8498}  + 1.0 EW_{\lambda  8542} + 0.6  EW_{\lambda 8662}$,
$V_{HB}$  being  a surface  gravity  correction  relative  to the  $V$
magnitude  of  the  horizontal   branch,  and  $V_{ave}$  the  average
(luminosity weighted) magnitude  of the stars in the  field.  We adopt
the value  of $V_{HB}=25.17$  for M31, measured  by \citet{holland96}.
The resulting combined spectra  are displayed in Figure~26, along with
the  metallicity measurement (on  the \citealt{carretta}  scale).  The
typical uncertainty is $\sim  0.2$~mags. By chance, the $\lambda 8498$
line in  the summed spectrum  in Field~16 falls  at the location  of a
night-sky line, so the two other Ca~II were used in that instance. For
comparison to the inner fields, on  the top panel of Figure~26 we also
display the  summed spectrum of  the disk fields D2--D5,  derived from
those stars within $2\sigma$ of the fit shown in Figure~10.

These metallicity measurements are  all very similar, identical within
the uncertainties, and agree reasonably  well with the position of the
targets  relative to  our  fiducial clusters  on the  color-magnitude
diagram.  This yields further evidence that the dominant population in
the  majority  of  the  fields  we surveyed  starting  from  the  disk
populations at a  radius of $15\kpc$ out to  $\sim 50\kpc$ are probing
different parts of the same global population, which has metallicity
of ${\rm [Fe/H] = -0.9}$.

\begin{figure}
\begin{center}
\includegraphics[angle=0,width=\hsize]{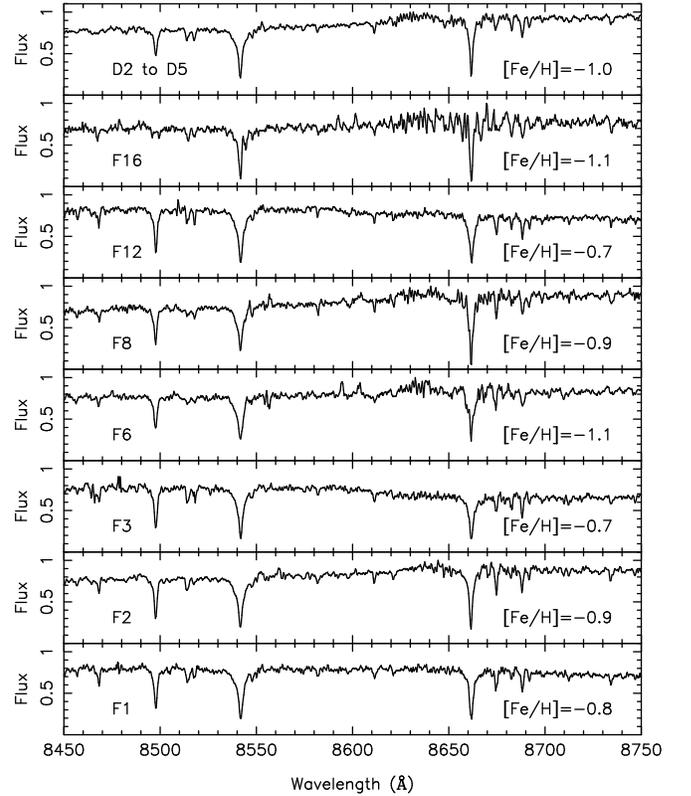}
\end{center}
\caption{Combined spectra  for all the  stars within $2\sigma$  of the
Gaussian  fit.   The  three   Ca~II  triplet  lines  at  8498.02\AA,
8542.09\AA, and 8662.14\AA\ are  the most obvious features of these
spectra.  The  field  name  is  indicated on  the  left,  and  derived
metallicities on the right of each panel.}
\end{figure}

\subsection{Scale-length}

A further useful constraint on the nature of this disk-like population
is its  density profile. We proceed  by measuring the  number $n_s$ of
stars  within $2\sigma$  of the  narrow peaks  shown in  Figure~10 and
Figures 16  to 23, and count  the number of observed  targets $n_o$ in
the CMD  selection window,  and also the  number of  available targets
$n_t$ within the  selection window that fall into  the DEIMOS field of
view.  The local density  in a field is estimated as $n_t  n_s / n_o -
n_b$, where  $n_b$ is the  background density computed from  fields at
the  edge  of  the survey.   The  top  panel  of Figure~27  shows  the
resulting  RGB  star density  profile,  under  the  assumption of  the
standard \citet{walterbos88}  inclination angle of $i  = 77\deg$.  The
profile shows an  exponential decline to  $R \sim 40\kpc$, at
which point there appears to be a break in the relation, with possibly
a flat  extension from  $\sim 40\kpc$ to  $\sim 70\kpc$.   Fitting the
data at $R < 40\kpc$, we measure an exponential scale-length of $h_R =
6.8 \pm  0.4 \kpc$.  This value  is somewhat higher than  the $5.1 \pm
0.1 \kpc$  scale-length (dotted  line) that we  fitted earlier  to the
major  axis star-counts. If  we instead  assume the  lower inclination
angle  calculated  in  \S2\  of  $i =  64.7\deg$  (lower  panel),  the
scale-length drops marginally to $h_R = 6.7 \pm 0.3 \kpc$.

\begin{figure}
\begin{center}
\includegraphics[angle=0,width=\hsize]{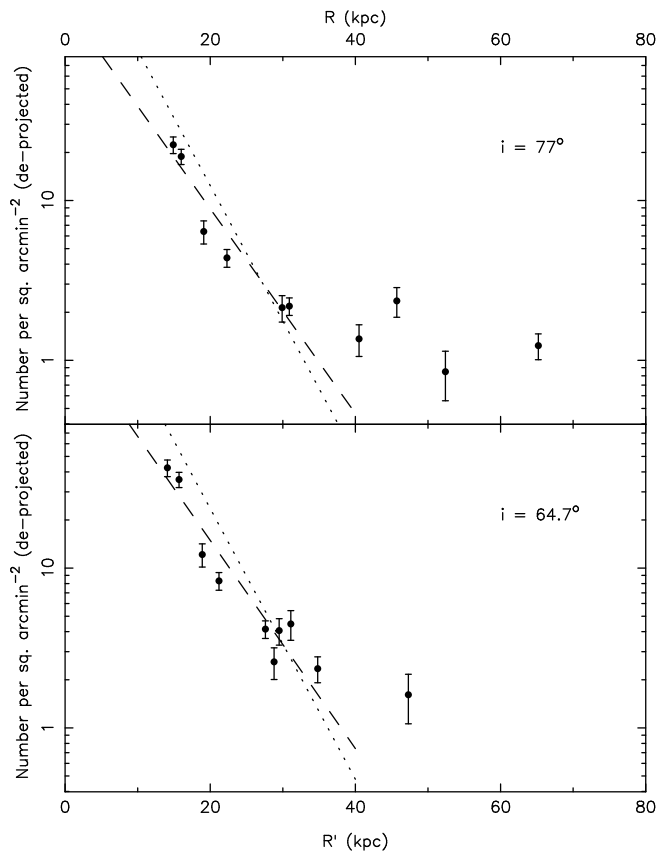}
\end{center}
\caption{The  density  profile  of  the extended  disk  population  is
displayed    in    the     top    panel,    isolated    by    counting
kinematically-confirmed  stars and  scaling by  the ratio  of possible
targets  to the  total  number of  stars  measured in  each field.   A
disk-plane inclination  of $i =  77\deg$ is assumed to  calculate both
the de-projected radius $R$  and the de-projected surface density. The
dashed line  shows an exponential  fit to the  data in the range  $R <
40\kpc$, which  implies a  scale-length of $h_R  = 6.8 \pm  0.4 \kpc$.
The dotted line reproduces the fit  to the outer major axis profile of
Figure~2 ($h_R = 5.1 \pm 0.1$).  Field F6 is not included in this plot
since  it  lies on  the  dwarf  galaxy  NGC~205, whose  stars  enhance
enormously the  number of  possible targets in  the DEIMOS  field. The
lower  panel shows  the  same data  for  $i =  64.7\deg$.  The  fitted
exponential function has  scale-length $h_R = 6.7 \pm  0.3 \kpc$.  The
dotted line  again reproduces the  fit of Figure~2, corrected  for the
difference in inclination angle.}
\end{figure}

These longer scale-lengths may mean that the population with disk-like
kinematics  that we  have uncovered  in  these fields  has a  slightly
different profile to the underlying exponential population revealed in
Figure~2, or it  may also mean that the inclination  of the outer disk
is lower than $i = 64.7\deg$.  However, it should be recalled that the
spectroscopic  fields preferentially targeted  overdensities detected
in the photometric survey, which  need not follow closely the medianed
major axis profile.   Another important property to bear  in mind when
interpreting  Figure~27  is that  the  velocity  distributions in  all
fields showed  narrow peaks at a  similar velocity lag.   There was no
indication  of  bimodal  distributions  with  a component  due  to  an
underlying smooth ``normal'' disk  plus an additional component due to
the sub-structure (with  the exception of field F2  which we discussed
in \S5.2).  If there were  an underlying normal disk population it has
identical kinematics to the substructure.

\subsection{Total luminosity and angular momentum}

In terms  of its mass fraction,  the outer disk is  a very significant
part of  the galaxy.   As we have  shown in Figure~2,  the exponential
decline  can  be followed  out  to  $40\kpc$  or $7.8$  scale-lengths.
Approximately 10\% of  the mass of the M31  disk therefore lies beyond
the  typical  $4$  scale-length  cutoff \citep{pohlen}.   This  region
contains an even larger fraction  of the angular momentum of the disk:
assuming a flat  rotation curve results in $\sim  30$\% of the angular
momentum of the disk residing beyond 4 scale-lengths.

\subsection{\ion{H}{1} content}

Recently, \citet{thilker04a} have presented  an analysis of Green Bank
Telescope  $\lambda 21\cm$  observations of  M31 and  its environment.
Their very high sensitivity of  $2.5 \times 10^{17} \cm^{-2}$ within a
beam-size of  $2\kpc$ allowed them to detect  many low-mass \ion{H}{1}
clouds  as  well  as  an  extended filamentary  component.  A  careful
inspection  of their Figure~1,  in which  the \ion{H}{1}  data-cube is
presented, reveals  a spatial overlap between  the \ion{H}{1} emission
and  our extended  disk  fields. Furthermore,  the  velocities of  the
\ion{H}{1} gas coincide  with the peaks of the  extended disk velocity
distributions found in the present study, though a detailed comparison
is  not  currently possible  as  the  velocity  intervals in  the  map
presented by \citet{thilker04a} are large ($34\kms$).

\subsection{Variations in stellar populations}

The color-magnitude diagrams presented  in Figures 16--23 show a very
similar  broad upper RGB,  indicating that  a similar  distribution of
stars  populates these  fields.   However, recent  work  by our  group
\citep{ferguson05} using very deep photometry from the Advanced Camera
for  Surveys (ACS)  instrument  on board  the  Hubble Space  Telescope
reveals significant  differences in  the morphology of  the horizontal
branch over the  inner part of the M31  ``halo''.  The surveyed fields
include small  areas of the fields F1,  F5, F9, F10, F11  and F12 from
the   present  study.   The   differences  in   the  horizontal-branch
populations  testify to the  existence of  inhomogenities in  the {\it
old} stellar populations in these fields.

\section{Discussion}

Our  radial  velocity  data  reveal  the  presence  of  an  unexpected
population, present in almost  all the surveyed fields from $R=15\kpc$
out to  $\sim 70\kpc$, that lags  the \ion{H}{1} by  at most $80\kms$,
and that  has a  dispersion of $\sim  20\kms$ to $\sim  40\kms$.  This
velocity dispersion is  large enough to rule out  the possibility that
these structures are bound dwarf  galaxies.  It should be kept in mind
that the circularly-rotating disk  model that we presented in Figure~7
is very simplistic, and  one should expect significant deviations from
it, especially  on fields  at large $y$  (minor axis  distance), where
small  changes in the  inclination of  the disk,  due for  instance to
warping, will cause large errors in the disk velocity lag. So the good
agreement in the mean velocity of  the disk-like peak in each field is
all the more  surprising. The direction of the effect  is such that if
the  disk is warped  so that  it is  seen less  edge-on, the  model of
Figure~7 will  overestimate the rotational velocity,  and a population
will appear to run ahead of circular motion.

The errors  in the model,  however, will not affect  significantly the
fitted $\sigma_v$ in  each field. The variations in  $\sigma_v$ in the
outer fields are  large, ranging between $\sigma_v =  15\kms$ in field
F3 and  $\sigma_v = 38\kms$ in  field F8.  Field F2  (with $\sigma_v =
35\kms$) mirrors field F3 about the major axis, so we can see that the
differences  in  velocity  dispersion  reflect  real  changes  between
fields, which are not just due to different projections of the $R$ and
$\phi$ components of the velocity ellipsoid onto the line of sight.

For the metallicity measurements we are unfortunately forced to co-add
the spectra in the velocity peaks to obtain combined spectra of useful
signal to noise.  This necessarily means that we  lose all information
on the  metallicity dispersion  of the populations.  Nevertheless, the
mean metallicity can  be measured reliably, and the  derived values in
all fields  are consistent with the  data from each  field being drawn
from a global population of identical metallicity.

All the DEIMOS  fields we present in this paper,  except for fields F5
and F7 were positioned on regions of high target density in M31, which
is  seen   as  the  dark  black   area  in  a   visual  inspection  of
Figure~1. This  corresponds to  a surface number  density of  stars of
approximately $10^4$ degree$^{-2}$ in the RGB selection window defined
in \S3. The  panoramic view of the INT  survey provides an invaluable
complement  to the  spectroscopic  results. Given  the very  similar
velocity  dispersions, disk  velocity  lags and  metallicities of  the
stars we have  identified, it is clear from  an inspection of Figure~1
that these populations can be united in a single morphological entity:
an   extended  disk-like   structure  which   has   rather  remarkable
well-defined fragmented sub-structures at its outer edge.

\subsection{Interpretation of velocity lag}

We  have  found a  population  that appears  to  lag  the velocity  of
circular orbits  in many fields, but  what does this  lag represent? A
simple answer may be that, in reality, the outer disk is less inclined
than we modeled.  However, some  fraction of the velocity lag must be
intrinsic.  In an equilibrium disk, a stellar population with non-zero
velocity  dispersion lags  the circular  speed; this  is known  as the
asymmetric  drift  (see,  \eg\  \citealt{BT}, their  \S4.2).   We  use
Eqn. 4-35 of \citet{BT} to estimate the expected asymmetric drift, and
adopt  their  simplifying assumptions  taken  to  calculate the  Solar
Neighborhood asymmetric drift  relation. Figure~28 shows the expected
asymmetric  drift as  a function  of  radius for  disk populations  of
dispersion  $\sigma_\phi  =  20\kms$  (dotted  line),  $\sigma_\phi  =
30\kms$ (continuous  line) and  $\sigma_\phi = 40\kms$  (dashed line).
These  estimates  show  that  a  significant velocity  lag  should  be
expected.

\begin{figure}
\begin{center}
\includegraphics[angle=270,width=\hsize]{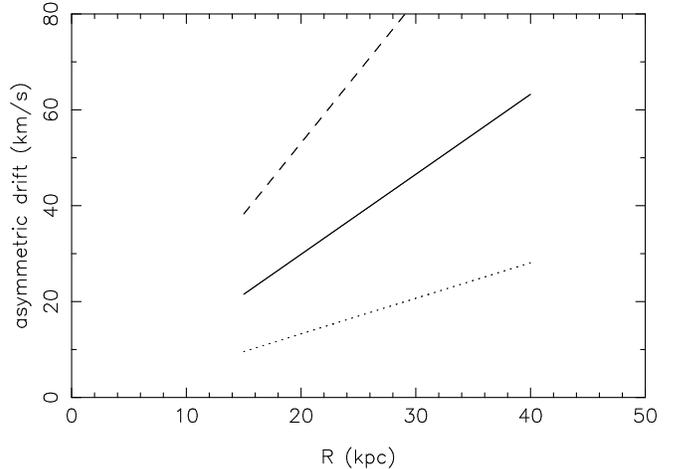}
\end{center}
\caption{The asymmetric drift of a disk population in equilibrium, for
three different values of the $\phi$-component of velocity dispersion:
$20\kms$  (dotted  line),  $30\kms$  (continuous  line)  and  $40\kms$
(dashed line). A scale-length  of $5.1\pm0.1\kpc$, as measured in \S2,
was adopted.}
\end{figure}

However, there are many  uncertainties that hinder a direct comparison
of these  asymmetric drift  estimates to the  data. First of  all, the
overdensities  targeted   in  the  outer  fields   are  evidently  not
equilibrium structures, so the asymmetric drift equation need not hold
precisely.   Apart from  the  uncertainty in  the  inclination of  the
galaxy,  there are  further  uncertainties in  the  projection of  the
velocity dispersion ellipsoid, as  we discussed in \S4.1.  A realistic
treatment of the asymmetric drift will require data that sample better
the extended disk region.

\subsection{Time-scale for dissolution}

With  typical  sizes  of  $\sim  5\kpc$ and  velocity  dispersions  of
$\sigma_v  \sim   30\kms$,  the   virial  mass  associated   with  the
sub-structures we  discuss in \S5\  would have to  be of the  order of
$\sim   10^9\msun$.    However,   the   total  luminosity   of   these
sub-structures is very  low, although this is a  difficult quantity to
measure  given   the  irregular  morphology  and   the  difficulty  of
identifying their  spatial limits.  For  the case of field  F16, where
the structure  is well-separated from  M31, the total luminosity  is a
mere $3\times  10^6\lsun$.  Thus if  these structures are  bound, they
require a ridiculous  mass to light ratio of  several hundred in Solar
units.  They  are instead most probably  unbound structures dissolving
in the global potential of the galaxy.

It is informative  to estimate the lifetime of  the structures if they
are indeed unbound. In this  case both the winding due to differential
rotation  and   the  velocity  dispersion  of  the   structure  set  a
dissolution time-scale.  In a potential  with a flat rotation curve of
velocity  $v_c$, an  unbound  structure at  galactocentric radius  $R$
becomes smeared out  by an extent equal to its initial  size in a time
$R/v_c$, i.e. in $170 \Myr$ at  $R=40\kpc$ for the case of M31 (taking
$v_c =  240\kms$).  Alternatively, if  we consider just the  effect of
the expansion of the stellar population due to its velocity dispersion
unconstrained  by self-gravity,  with a  velocity dispersion  of $\sim
30\kms$,  a structure  of size  $5\kpc$ will  double in  size  in $170
\Myr$,  fortuitously the  same  time estimated  before.  These  simple
arguments give an order-of-magnitude estimate of the the survival time
of  the   structures,  and  strongly  suggest  that   they  cannot  be
long-lived.

\subsection{Age of population}

\citet{brown} present  extremely deep  photometry with the  ACS camera
for a  field on  the minor axis  of M31.  The location of  this field,
indicated in Figure~4 with a  small square, lies at a projected radius
of only $y = -  11.4\kpc$. The corresponding de-projected radius is $R
=  53\kpc$,  beyond  the  break  in the  surface  density  profile  of
Figure~27. However, a  tiny $5^\circ$ warp in the  extended disk would
bring the field into the $R  < 40\kpc$ region. In their data, Brown et
al.  find evidence  for a  30\%  component of  intermediate age  stars
($6$--$8\Gyr$)  with metallicity  ${\rm [Fe/H]  > -0.5}$,  though they
claim  a similar-quality  fit for  a population  of  metallicity ${\rm
[Fe/H] >  -1.0}$. Given  that this is  the youngest component  in that
field,  and  has  comparable  metallicity,  it is  possible  that  the
intermediate-age population  whose age they constrain  is the extended
disk. However, a definitive measure  of the age of the population will
require  resolving the main-sequence  turnoff in  fields where  we are
sure the the extended disk population is present.

\subsection{Formation scenarios}

How  did  the  extended   disk  form?  Several  possibilities  present
themselves.

\begin{itemize}

\item  {\bf Formation  by  the accretion  of  many small  sub-galactic
structures.}   This   scenario  would  explain   naturally  the  lumpy
sub-structure seen  at the edge of  the extended disk, as  well as the
differences in the  velocity dispersion and the variations  in the old
stellar populations between fields.  However, the chemical homogeneity
of the  material argues against this possibility,  since the different
accreted   galaxies   would  likely   have   a   range  of   different
metallicities,  depending on  their mass  and  star-formation history.
Admittedly, the  metallicity measurements  we present above  give only
the  mean  of the  population,  and in  the  central  regions of  M31,
considerable  mixing  must  have  taken  place.  Nevertheless,  it  is
surprising  to find  the same  mean  abundance even  in the  outermost
sub-structures. A further problem  with this scenario is the disk-like
kinematics. A  substantial mass is required for  dynamical friction to
affect the  orbit of a galaxy satellite  ($> 10^8\msun$, \citealt{BT})
which has to become circularized {\it before} being disrupted in order
to give rise  to a final structure with  disk-like kinematics. Another
problem is the disk-like  distribution of debris.  Though accretion in
the  disk plane  does appear  to be  favored  \citep{penarrubia}, the
accretion of a  large number of halo objects is  expected to give rise
to a distribution  closer to that of a spheroidal  stellar halo than a
disk. Even the thick disk component in the \cite{abadi} simulation, in
which $\sim 50$\% is formed by accretion, is a very thick component of
axis ratios $\sim 1:3$ and scale-length of $4.5\kpc$ (our measurements
from their Figures 1 and 3). Forming the extended disk observed in M31
in this way will certainly be a challenge for modelers.
\item {\bf  Formation from stars  accreted in a single  large merger.}
Here we  envisage the  assimilation of a  whole M33-sized galaxy  in a
single merging  event.  Accreting 10\% of  the total mass  of a galaxy
counts as a significant merger, affecting the morphology of the larger
galaxy. However,  given that  the bulk of  the interaction  took place
long ago, possibly $6$--$8\Gyr$ in  the past (corresponding to $z \sim
1$), the progenitor  of M31 would have been  considerably less massive
at  that  time \citep{hammer},  and  the  resulting interaction  would
likely be termed  a ``major merger'' after which  Andromeda would have
been globally  reshaped.  Thin disks  are easily heated up  by mergers
\citep{toth, velazquez}, so this  scenario makes a testable prediction
that stars  in the  inner disk  of M31 that  formed before  the merger
should have a much larger scale-height than the younger thin disk. The
advantage of  this proposition is  that it accounts naturally  for the
uniform  mean abundances  measured  in the  different  fields.  It  is
questionable that such a merger  could have given rise to a population
with disk-like  morphology and  kinematics, since typical  disk galaxy
collisions     result     in     elliptical    end-products     (e.g.,
\citealt{hernquist92}).  However,  new simulations by \citet{springel}
cast doubt  on the inevitability  of forming elliptical  galaxies from
such collisions,  showing that  disk galaxies may  form even  in equal
mass mergers if the  progenitors are gas dominated.  Further numerical
experiments are  required to establish  whether a disk as  extended as
the one seen around M31 can be formed in this way.
\item {\bf Formation induced by  a single large merger.} This scenario
would be  similar to the previous  one, but the  stellar population of
the  accreted  galaxy would  end  up  populating  the bulge  or  other
spheroidal  component of  M31.  This  is perhaps  consistent  with the
presence of a double nucleus in that galaxy \citep{lauer}. However the
interaction itself  could have  triggered star-formation on  a massive
scale in the pre-existing gaseous  disk of M31, with gas accreted from
the satellite contributing to  the enrichment.  This possibility could
explain the  disk-like structure and kinematics, and  has the testable
prediction  that its  constituent stars  should be  no older  than the
interaction  itself.   The  homogeneity  in mean  abundance  over  the
structure  could be  explained  by star-formation  being limited  very
rapidly  before chemical enrichment  could take  place.  Such  a rapid
formation of  the disk in this way  is similar to what  is deduced for
$\sim 1  {\rm mJy}$ sub-mm sources  at $z \sim  2$ with star-formation
rates  of $\sim 100  \msunyr$ \citep{chapman03,  greve}. Nevertheless,
this scenario is also problematic, as it is hard to reconcile with the
variations in the velocity dispersion between fields.
\end{itemize}

A concern with the last  two scenarios is that the sub-structures that
we  see on  the outskirts  of the  extended disk  now  are short-lived
features that  will soon  disappear ($<< 1  \Gyr$). This means  that we
would be observing  the system at a special time  at the very tail-end
of the  merging process.  A further  problem is the  constraint on the
size of  a past major merger set  by the existence of  a population of
thin disk globular  clusters that has recently been  discovered in M31
by \citet{morrison04}. Those  authors infer that no more  than 10\% of
the mass of the disk could have been accreted in a single merger since
the formation of  the clusters. The time of  formation of the clusters
is currently  poorly constrained, though  they find no reason  why the
clusters should be significantly younger  than those of the Milky Way.
Measuring   the  ages   of  these   objects  will   help   greatly  in
distinguishing between the above scenarios.

\subsection{Possible connection to NGC~205?}

As we noted  in \S5.4, if stars become unbound  from NGC~205, we would
observe them in field F6  with kinematic properties very close to that
of disk material. Likewise,  if \citet{mcconnachie04b} were correct in
identifying  the arc-like structure  to the  North of  M31 as  a tidal
stream from NGC~205, this tidal  debris, which is located in fields F8
and F9, would also have disk-like kinematics.

This raises the possibility that  NGC~205 is connected to the extended
disk  structure, either as  the direct  supplier of  stars, or  as the
remnant of the  ancient structure which created the  extended disk. It
is interesting  to note in this  context that the  mean metallicity of
NGC~205   has  been   measured  to   be  ${\rm   [Fe/H]   \sim  -0.9}$
\citep{mould}, identical  to the mean  spectroscopic measurements from
the fields presented in this survey.

A counter-argument however, is  that recent estimates for the distance
to NGC~205 indicate that it is behind M31, being at a distance of $824
\pm   27  \kpc$   \citep{mcconnachie05},  compared   to   $785\pm25  -
40.6\sin(i) \kpc$  for the  M31 disk at  the location of  NGC~205. The
difference  in  distance  is  therefore $79\pm  37\kpc$,  little  over
$2\sigma$,  although  we   are  probably  grossly  overestimating  the
uncertainties, since  a differential distance  measurement is expected
to be much more accurate.

However, this  need not rule out  the connection with  NGC~205. If the
dwarf galaxy were  indeed $\sim 80\kpc$ beyond the  foreground disk on
the  near side of  M31, it  would be  $\sim 40\kpc$  behind M31,  at a
location $\sim 17\kpc$ above the plane  of the disk on the far side of
that galaxy.

\subsection{Comparison to \citet{hurleykeller}}

\citet{hurleykeller} have  presented an analysis of  the kinematics of
135 planetary nebulae in the disk and inner halo of M31. They find the
surprising result that a standard  model for the thin disk, thick disk
and bulge cannot  reproduce the observed PNe kinematics  at major axis
distances  of $15\kpc$  to $25\kpc$.   In order  not to  introduce the
presence  of an  additional component,  they invoke  Occam's  razor to
alter the parameters of the bulge, deducing that the bulge must rotate
rapidly at  large distances. We re-assess  their data in  the light of
the discovery  of the  additional extended disk  population. Figure~29
shows the positions of the 135 PNe in their survey. Stars close to the
centre of M31 are a complex  mix of populations, so we investigate the
sub-sample with $|x| > 8 \kpc$, which are displayed as filled circles.
The distribution of  velocity lags is shown in  Figure~30. We fit this
distribution  for  $v_{lag}  >  -150\kms$ using  a  maximum-likelihood
algorithm,  to  account  for  the velocity  measurement  uncertainties
quoted by \citet{hurleykeller} as well as the disk model uncertainties
(from Figure~8).  The resulting Gaussian model has $\overline{v}_{lag}
=  -52\kms$ and  $\sigma_v =  44  \kms$, very  similar to  the fit  in
Figure~10.   Further analysis is  necessary to  establish conclusively
whether the existence  of the extended disk alleviates  the need for a
rapidly rotating outer bulge as suggested by \citet{hurleykeller}.

\begin{figure}
\begin{center}
\includegraphics[angle=270,width=\hsize]{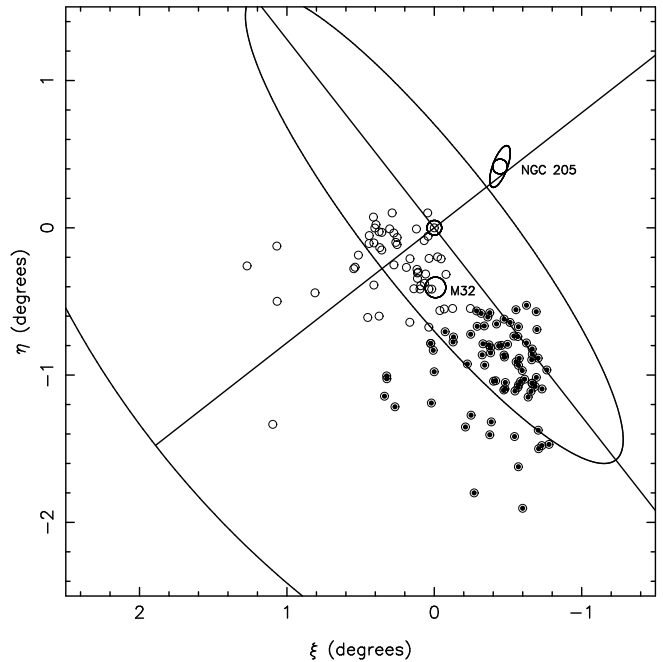}
\end{center}
\caption{The  location  of  the  PNe  from  \citet{hurleykeller}  with
respect  to M31;  the  lines and  ellipses  from Figures~1  and 4  are
reproduced to  show the  scale. The small  open circles show  the full
sample of 135 PNe, while our chosen sub-sample of 77 objects with $|x|
> 8\kpc$ are indicated with filled circles.}
\end{figure}

\begin{figure}
\begin{center}
\includegraphics[angle=270,width=\hsize]{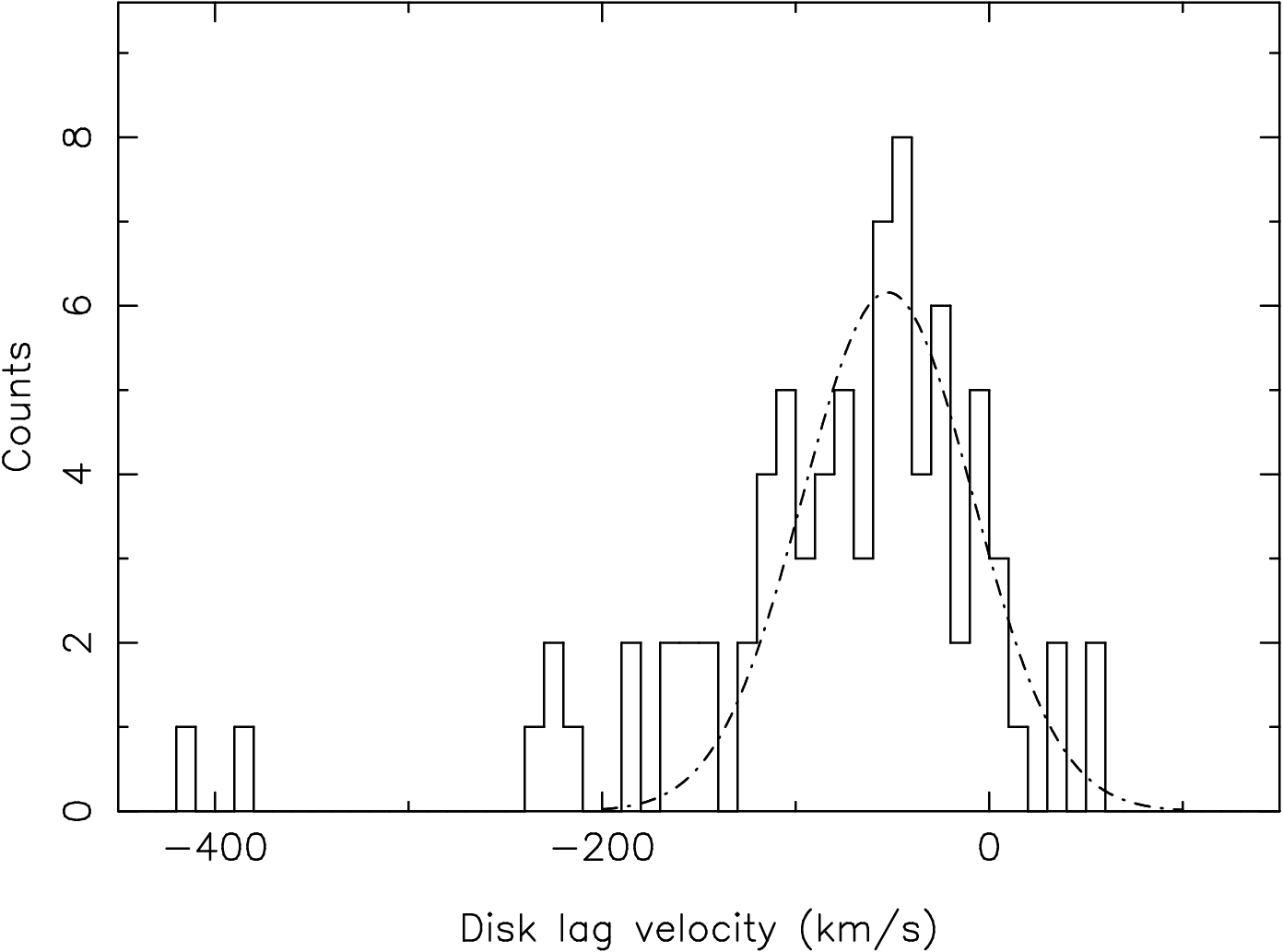}
\end{center}
\caption{The   distribution   of    disk   lag   velocities   of   the
\citet{hurleykeller}  PNe  with $|x|  >  8\kpc$. A  maximum-likelihood
Gaussian  fit  to the  data  in the  interval  $-150\kms  < v_{lag}  <
100\kms$  is  also  shown,  for  which  both  the  estimated  velocity
uncertainties from  \citet{hurleykeller} as well  as the uncertainties
in  the disk  rotation  model  (from Figure~8)  have  been taken  into
account.  We fit a Gaussian mean velocity lag of $\overline{v}_{lag} =
-52\kms$     and      dispersion     $\sigma_v     =      44     \kms$
($\overline{v^\prime}_{lag}   =  -44\kms$,  $\sigma_{v^\prime}   =  46
\kms$). The same low  dispersion disk-like population as discovered in
our DEIMOS fields is also present here.}
\end{figure}

\subsection{Globular clusters}

A population of  massive globular clusters has recently  been found in
the disk of M31 \citep{fusi-pecci}. These objects are scattered around
the outskirts  of the  inner disk at  distances of $\sim  20\kpc$, and
have very blue colors, with  color indices indicating a young age of
$\sim 2\Gyr$. The population  is also very significant, accounting for
$\simgt  15$\% of  the globular  cluster system  of that  galaxy. This
population  of young clusters  rotates rapidly,  possessing kinematics
similar to that of the  disk.  A further piece of relevant information
is  provided   by  the  recent  discovery  of   the  globular  cluster
Bologna~514 \citep{huxor03,  galleti}, which is situated  close to the
South-western major axis  of M31 at a projected  distance of $55\kpc$,
almost exactly  on the outer ellipse  marked in Figures~1  and 4.  The
rotation model of Figure~7 gives  a disk-lag velocity of this globular
cluster of $v_{lag} = -70\kms$ (${v^\prime}_{lag} = -53\kms$).

Though it  is tempting to  speculate on a possible  connection between
these globular  clusters and the  extended disk, the  metallicities of
the clusters appear to be much lower than the ${[\rm Fe/H] = -0.9}$ we
measure for the structures in  this study.  However, as pointed out by
\citet{fusi-pecci}, the metallicities of these globular clusters could
be  higher  if  they  are   of  younger  age  than  the  typical  halo
cluster. Further observations are needed to establish the age of these
M31 disk clusters, and their relation to the extended disk.

\subsection{A connection with the thick disk?}

Several  edge-on  galaxies have  been  found  to  possess a  flattened
component  with  a  much   larger  scale-height  than  the  thin  disk
\citep{dalcanton02}  probably  indicating  the  presence  of  a  thick
disk. Recently, two of the sample of \citet{dalcanton02} were observed
spectroscopically by  \citet{yoachim}, who found the thick  disk to be
rotating very slowing  in one case and counter-rotating  in the other.
This  result shows  that  thick disks  can  have completely  different
kinematics from the  thin disk, although this is not  the case for the
Milky Way.  The fact that the extended  disk we find around  M31 has a
low  velocity dispersion,  and  follows circular  orbits closely  thus
behaving  kinematically like a  thin disk,  as well  as having  a very
similar scale-length to the thin disk, suggests a close link with that
component,  and argues  against  association with  the  thick disk  of
Andromeda.   It should be  noted, however,  that it  has not  yet been
shown that  Andromeda does indeed  possess a thick disk.   The present
dataset may allow  us to detect the thick disk, but  we will return to
this issue in a future contribution.

\subsection{Extended disks in other galaxies?}

Our group has also been attempting to unravel the formation history of
the  Milky Way \citep{ibata01a,  ibata02}, with  similar goals  to our
study  of  M31. Following  up  on the  discovery  of  a peculiar  halo
structure in the SDSS \citep{newberg}, we showed that the Milky Way is
enveloped  in  a Ring-like  distribution  of  old  stars that  lie  at
Galactocentric distances of $\sim 20$~kpc, well beyond the edge of the
Galactic disk  \citep{ibata03}.  A subsequent  analysis of asymmetries
in the 2MASS survey revealed a strong enhancement of stars at the edge
of the Galactic disk, which we interpreted as an accreted dwarf galaxy
fragment  (\citealt{martin04a, martin04b,  bellazzini}, see 
also \citealt{rocha-pinto05}); a
connection  with  the  ring-like  structure is  possible,  though  not
proven.   However,  this ``Galactic  Ring''  or ``Galactic  Anticenter
Stellar Structure''  \citep{crane} is not the  only stellar population
with disk-like  kinematics that lies beyond  the rim of  the Milky Way
thin-disk: recently  \citet{majewski04} and \citet{rocha-pinto04} have
uncovered a further  stellar structure located at a  distance of $\sim
40$~kpc  and which  spans a  huge angle  on the  sky.   These findings
indicate that the  outer regions of the Galactic  plane have a similar
messy flattened structure to what is observed in M31 (Figure~1).

A possible  young counterpart to  the ``extended disks'' in  the Milky
Way  and M31  has  been seen  in  some more  distant galaxies.   GALEX
observations  of  M83  \citep{thilker04b}  have revealed  an  extended
population of UV  sources (OB stars) beyond the  inner disk (though at
much  smaller radius  than  the  structure we  detect  in M31),  while
several galaxies are known to exhibit \ion{H}{2} regions to extremely
large radii \citep{ferguson98}.  Thus  it appears that there may exist
an   additional    component   to   spiral    galaxies,   a   putative
``extended-disk''.    The  implications   for  galaxy   formation  are
profound:  the inhomogenous spatial  distribution and  moderately high
metallicity suggests  an origin  in multiple accretions,  yet accreted
fragments, which are the building blocks of galaxies, do not give rise
to these huge extended disks in current simulations.

\section{Conclusions}

We  have presented  an analysis  of a  large survey  of  the Andromeda
galaxy  with the  Keck/DEIMOS spectrograph  based on  targets selected
from  a  panoramic survey  with  the INT  Wide  Field  Camera. A  huge
rotating  structure  that  follows  an exponential  density  law  with
scale-length  $5.1 \pm  0.1 \kpc$  is detected  out to  a de-projected
distance  of $\sim  40\kpc$, with  sub-structures continuing  out even
further in  radius.  All of  the photometric sub-structures  that were
targeted for spectroscopic follow-up participate in the same disk-like
movement.  The velocity dispersion  is low, typically $30\kms$.  These
properties  indicate that the  population is  a huge  extended stellar
disk.  The color  dispersion on the RGB of stars  that belong to this
structure is broad, implying that a wide metallicity range is present.
Spectroscopic measurements of the mean metallicity give ${\rm [Fe/H] =
-0.9 \pm 0.2}$.   No significant radial trend in  either kinematics or
metallicity is found,  although deep ACS photometry in  several of the
DEIMOS fields  reveals variations in the  horizontal branch morphology
over the  galaxy, indicating that  the stellar population of  the most
ancient stars differ between fields.   We estimate that $\sim 10$\% of
the luminous mass of the  Andromeda disk resides in this extended disk
component,  beyond  the  $4$  scale-length  cutoff  observed  in  most
galaxies.  However,  this region  may account for  $\sim 30$\%  of the
total angular momentum of the galaxy.

The spatial sub-structures that form the outer edge of this system are
almost certainly no longer gravitationally bound and will disappear as
spatially  coherent  entities  within  $\sim 200\Myr$.   However,  the
existence of  these sub-structures indicates a  formation mechanism by
mergers, and implies that the extended disk is still forming, although
we  seem to be  observing it  at the  very tail  end of  its formation
period.  It  is unclear  at this point  whether the extended  disk was
formed by multiple accretion events or by a single large event, though
the latter possibility appears more plausible given the present data.

Many  questions  remain  to  be  answered.  Is  the  extended  disk  a
continuation of the thin disk, or  is it a separate component? Was the
merger, or were the mergers, that formed the extended disk responsible
for the formation of the  $R^{1/4}$-law halo of Andromeda?  Why is the
extended  disk closely  aligned  with the  thin  disk? Why  is it  not
extremely warped?

The surface brightness of  this structure, 27-31~mag/arcsec$^2$ in its
outer regions, places  it beyond what would be  classified as ``disk''
in more  distant systems. Though such faint  surface brightness levels
are extremely challenging to observe, by resolving individual stars it
will  be  possible  to  ascertain  whether  the  extended  disk  is  a
particularity  of M31  or whether  similar components  are  present in
other galaxies. Further work is also  needed in the Milky Way to study
the distant  stellar populations  hidden by foreground  populations in
the Galactic  plane, and  to understand the  nature of  the structures
recently  unveiled  at  the  edge  of the  Galactic  disk  and  beyond
\citep{newberg,    ibata03,    martin04a,    martin04b,    majewski04,
rocha-pinto04, rocha-pinto05}, and their relation to the vast extended
disk in the Andromeda galaxy.

\section*{Acknowledgements}

RI would  like to thank Fran{\c  c}oise Combes, and  Bernd Vollmer for
very helpful comments and  suggestions.  SCC acknowledges support from
a NASA.  GFL acknowledges support through the ARC DP0343508.


\begin{thebibliography}{}
\bibitem[Abadi et al.(2003)]{abadi}
        Abadi M., Navarro J., Steinmetz M. \& Eke V. 2003, \apj\ 597, 21
%
\bibitem[Avila-Reese, Firmani \& Hern\'andez(2002)]{avila-reese}
        Avila-Reese, V., Firmani, C., Hern\'andez, X. 2001, in 
        New Quests in Stellar Astrophysics: 
        The Link between Stars and Cosmology, 
        eds. M. Chavez, A. Bressan, A. Buzzoni, \& D. Mayya (Dordrecht: Kluwer)
%
%
\bibitem[Bellazzini et al.(2004)]{bellazzini}
        Bellazzini, M., Ibata, R., Monaco, L., Martin, N., Irwin, M., 
        Lewis, G. 2004, \mnras\ 354, 1263
%
%
\bibitem[Binney \& Tremaine(1987)]{BT}
        Binney, J. \& Tremaine, S. 1987, Galactic Dynamics, Princeton University
        Press, Princeton
%
\bibitem[Bland-Hawthorn et al.(2005)]{bland-hawthorn}
        Bland-Hawthorn, J., Vlaji\'c, M., Freeman, K., Draine,
        B. 2005, \apj\ in press, astro-ph/0503488
%
\bibitem[Brinks \& Burton(1984)]{brinks}
        Brinks, E., Burton, W. 1984, A\&A 141, 195
%
\bibitem[Brown et al.(2003)]{brown}
        Brown, T., Ferguson, H., Smith, E., Kimble, R., Sweigart, A., 
        Renzini, A., Rich, M., VandenBerg, Don A., 2003, \apj\ 592, 17L
%
\bibitem[Carretta \& Gratton(1997)]{carretta}
        Carretta, E., Gratton, R. 1997, A\&AS 121, 95
%
\bibitem[Chapman et al.(2003)]{chapman03}
        Chapman, S., Blain, A., Ivison, R., Smail, I. 2003, Nature 422, 695
%
\bibitem[Chiba \& Beers(2000)]{chiba00}
        Chiba, M., Beers, T. 2000, \aj\ 119, 2843
%
\bibitem[Crane et al.(2003)]{crane}
        Crane, J., Majewski, S., Rocha-Pinto, H., Frinchaboy, P., et al. 2003, ApJ 594, 119
%
\bibitem[Dalcanton \& Bernstein(2002)]{dalcanton02}
        Dalcanton J., Bernstein R., 2002, \aj\ 124, 1328
%
\bibitem[Davis et al.(2003)]{davis03}
        Davis, M., et al. 2003, Proc. SPIE, 4834, 161 
%
\bibitem[Dehnen \& Binney(1998)]{dehnen98}
        Dehnen, W. \& Binney, J., 1998, \mnras\ 298, 387
%
\bibitem[Durrell, Harris \& Pritchet(2001)]{durrell01}
        Durrell, P., Harris, W., Pritchet, C. 2001, \aj\ 121, 2557
%
\bibitem[Durrell, Harris \& Pritchet(2004)]{durrell04}
        Durrell, P., Harris, W., Pritchet, C. 2004, \aj\ 128, 260
%
\bibitem[Evans et al.(2000)]{evans00b}
        Evans, N., Wilkinson, M., Guhathakurta, P., Grebel, E., Vogt, S., 2000, \apj\ 540, 9L
%
\bibitem[Ferguson et al.(1998)]{ferguson98}
        Ferguson, A., Wyse, R., Gallagher, J., Hunter, D. 1998, \apj\
        506, 19
%
\bibitem[Ferguson et al.(2002)]{ferguson02}
        Ferguson, A., Irwin, M., Ibata, R., Lewis, G., Tanvir,
        N., 2002, \aj\ 124, 1452
%
\bibitem[Ferguson et al.(2005)]{ferguson05}
        Ferguson, A., Johnson, R., Faria, D., Irwin, M., Ibata, R.,
        Johnston, K., Lewis, G., Tanvir,
        N. 2005, \apj\ 622, 109L
%
\bibitem[Fusi Pecci et al.(2005)]{fusi-pecci}
        Fusi Pecci, F., Bellazzini, M., Buzzoni, A., De Simone, E., 
        Federici, L., Galleti, S. 2005, astro-ph/0503188
%
\bibitem[Galleti et al.(2005)]{galleti}
        Galleti, S., Bellazzini, M., Federici, L., Fusi Pecci,
        F. 2005, astro-ph/0503206
%
\bibitem[Gilmore \& Reid(1983)]{gilmore83}
        Gilmore, G., Reid, N. 1983, \mnras\ 202, 1025
%
\bibitem[Greve et al.(2005)]{greve}
        Greve, T., Bertoldi, F., Smail, I., Neri, R., Chapman, S., Blain, A.,
        Ivison, R., Genzel, R., Omont, A., Cox, P., Tacconi, L., Kneib, J.-P.
        2005, \mnras\ in press, astro-ph/0503055
%
\bibitem[Hammer et al.(2005)]{hammer}
        Hammer,F., Flores, H., Elbaz, D., Zheng, X., Liang, Y.,
        Cesarsky, C. 2005, \aap\ 430, 115
%
\bibitem[Hernquist(1992)]{hernquist92}
        Hernquist, L. 1992, \apj\ 400, 460
%
\bibitem[Holland(1996)]{holland96}
        Holland, S., Fahlman, G.\ G., Richer, H.\ B., 1996, AJ, 112, 1035
%
\bibitem[Hurley-Keller et al.(2004)]{hurleykeller}
        Hurley-Keller, D., Morrison, H., Harding, P., Jacoby, G. 2004,
        \apj\ 616, 804
%
\bibitem[Huxor et al.(2003)]{huxor03}
        Huxor, A. Tanvir, N., Irwin, M., Ferguson, A., Ibata, R., 
        Lewis, G., Bridges, T. 2003, in 
        Satellites and Tidal Tails, eds.  F. Prada, D. Martinez-Delgado, 
        T. Mahoney, ASP vol. 327
%
\bibitem[Huxor et al.(2005)]{huxor05}
        Huxor, A. Tanvir, N., Irwin, M., Ibata, R., Collett, J.,
        Ferguson, A., Bridges, T., Lewis, G. 2005, astro-ph/0412223
%
\bibitem[Ibata et al.(2001a)]{ibata01a}
        Ibata R., Lewis G., Irwin M., Totten E. \& Quinn T., 2001a, \apj\ 551, 294
%
\bibitem[Ibata et al.(2001b)]{ibata01b}
        Ibata, R., Irwin, M., Lewis, G., Ferguson, A., Tanvir,
        N., 2001b, Nature 412, 49
%
\bibitem[Ibata et al.(2002)]{ibata02}
        Ibata R., Lewis G., Irwin M. \& Cambr\'esy L., 2002, \mnras\ 332, 921
%
\bibitem[Ibata et al.(2003)]{ibata03}
        Ibata, R., Irwin, M., Lewis, G., Ferguson, A., Tanvir,
        N. 2003, \mnras\ 340, 21
%
\bibitem[Ibata et al.(2004)]{ibata04}
        Ibata, R., Chapman, S., Ferguson, A., Irwin, M., Lewis, G. 2004, \mnras\ 351, 117
%
\bibitem[Irwin \& Trimble(1984)]{irwin84}
        Irwin, M., Trimble, V. 1984, \aj\ 89, 83
%
\bibitem[Katz \& Gunn(1991)]{katz91}
        Katz, N. \& Gunn, J. 1991, \apj\ 377, 365
%
\bibitem[Kennicutt(1989)]{kennicutt}
        Kennicutt, R. 1989, \apj\ 344, 685
%
\bibitem[Klypin et al.(1999)]{klypin99}
        Klypin, A., Kravtsov, A., Valenzuela, O., Prada, F. 1999,
        {\apj} 522, 82
%
\bibitem[Klypin, Zhao \& Somerville(2002)]{klypin02}
        Klypin, A., Zhao, H.-S., Somerville, R., 2002, \apj\ 573, 597
%
\bibitem[van der Kruit \& Searle(1981)]{kruit}
        van der Kruit, P., Searle, L. 1981, A\&A 95, 105
%
\bibitem[Lauer et al.(1993)]{lauer}
        Lauer, T., Faber, S., Groth, E., Shaya, E., 
        Campbell, B., Code, A., Currie, D., Baum, W.,
        Ewald, S., Hester, J.,  Holtzman, J., Kristian, J.,
        Light, R., Ligynds, C.,  O'Neil, E., 
        Westphal, J. 1993, \aj\ 106, 1436
%
\bibitem[Loinard, Allen \& Lequeux(1995)]{loinard}
        Loinard, L., Allen, R., Lequeux, J. 1995, A\&A 301, 68
%
\bibitem[Martin et al.(2004a)]{martin04a}
        Martin, N. F., Ibata, R., Bellazzini, M., Irwin, M., Lewis, G.,
        Dehnen, W. 2004a, \mnras\ 348, 12
%
\bibitem[Martin et al.(2004b)]{martin04b}
        Martin, N., Ibata, R., Conn, B., Lewis, G., Bellazzini, M., 
        Irwin, M. 2004b, \mnras\  355, 33
%
\bibitem[Majewski et al.(2004)]{majewski04}
        Majewski S., et al. 2004, \apj\ 615, 738
%
\bibitem[McConnachie et al.(2004a)]{mcconnachie04a}
        McConnachie, A., {Irwin}, M., Ferguson, A., {Ibata}, R., 
        {Lewis}, G., Tanvir, N., 2004a, \mnras\ 350, 243
%
\bibitem[McConnachie et al.(2004b)]{mcconnachie04b}
        McConnachie, A., {Irwin}, M.,
        {Lewis}, G.,  {Ibata}, R., Chapman, S.,  
        Ferguson, A., Tanvir, N. 2004b, \mnras\ 351, 94
%
\bibitem[McConnachie et al.(2005)]{mcconnachie05}
        McConnachie, A., Irwin, M., Ibata, R., Lewis, G.
        Ferguson, A., Tanvir, N. 2005, MNRAS, astro-ph/0410489
%
\bibitem[Moore et al.(1999)]{moore99}
        Moore, B., Ghigna, S., Governato, F., Lake, G., Quinn, T., Stadel,
        J., Tozzi, P. 1999, {\apjl} 524, L19
%
\bibitem[Morrison et al.(2004)]{morrison04}
        Morrison, H., Harding, P., Perrett, K., Hurley-Keller, D. 2004,
        \apj\ 603, 87
%
\bibitem[Mould, Kristian \& Da Costa(1984)]{mould}
        Mould, J., Kristian, J., Da Costa, G. 1984, \apj\ 278, 575
%
\bibitem[Newberg et al.(2002)]{newberg}
        Newberg, H., et al. 2002, \apj\ 569, 245
%
\bibitem[Pe\~narrubia, Kroupa \& Boily(2002)]{penarrubia}
        Pe\~narrubia, J., Kroupa, P., Boily, C. 2002, \mnras\ 333, 779
%
\bibitem[Pohlen, Dettmar \& L\"utticke(2000)]{pohlen}
        Pohlen, M., Dettmar, R., L\"utticke, R. 2000, A\&A 357, 1
%
\bibitem[Reitzel \& Guhathakurta(2002)]{reitzel02}
        Reitzel, D., Guhathakurta, P. 2002, \aj\ 124, 234
%
\bibitem[Reitzel, Guhathakurta \& Rich(2004)]{reitzel04}
        Reitzel, D., Guhathakurta, P., Rich, M. 2004, \aj\ 127, 2133
%
\bibitem[Robin et al.(2004)]{robin}
        Robin, A., Reyl\'e, C.,  Derri\`ere, S., Picaud, S. 2004, \aap\
        416, 157
%
\bibitem[Rocha-Pinto et al.(2004)]{rocha-pinto04}
        Rocha-Pinto, H., Majewski, S., Skrutskie, M., Crane, J.,
        Patterson, R. 2004, \apj\ 615, 732
%
%
\bibitem[Rocha-Pinto et al.(2005)]{rocha-pinto05}
        Rocha-Pinto, H., Majewski, S., Skrutskie, M., Patterson, R. 2005, astro-ph/0504122
%
\bibitem[Ruphy et al.(1996)]{ruphy}
        Ruphy, S., Robin, A., Epchtein, N., Copet, E., Bertin, E.,
        Fouque, P.,  Guglielmo, F. 1996, \aap\ .313L..21R
%
\bibitem[Rutledge, Hesser \& Stetson(1997)]{rutledge}
        Rutledge, G., Hesser, J., Stetson, P. 1997, \pasp\ 109, 883
%
\bibitem[Schaye(2004)]{schaye}
        Schaye, J. 2004, \apj\ 609, 667
%
\bibitem[Schlegel et al.(1998)]{schlegel}
        Schlegel, D., Finkbeiner, D., Davis, M. 1998, {\apj} 500, 525
%
\bibitem[Springel \& Hernquist(2005)]{springel}
        Springel, V., Hernquist, L. 2005, \apj\ 622, L9
%
\bibitem[Steinmetz \& M\"uller(1995)]{steinmetz}
        Steinmetz, M. \& M\"uller, E. 1995, \mnras\ 276, 549
%
\bibitem[Summers(1993)]{summers}
        Summers, F. 1993, Ph.D. Thesis, University of California, Berkeley
%
\bibitem[Thilker et al.(2004a)]{thilker04a}
        Thilker, D., Braun, R., Walterbos, R., Corbelli, E. Lockman, F.,
        Murphy, E., Maddalena, R. 2004a, \apj\ 601, 39L
%
\bibitem[Thilker et al.(2004b)]{thilker04b}
        Thilker, D., et al. 2004b, astro-ph/0411306
%
\bibitem[T\'oth \& Ostriker(1992)]{toth}
        T\'oth, G., Ostriker, J. 1992, \apj\ 389, 5
%
\bibitem[Velazquez \& White(1999)]{velazquez}
        Velazquez, H., White, S. 1999, \mnras\ 304, 254
%
\bibitem[Walterbos \& Kennicutt(1987)]{walterbos87}
        Walterbos, R., Kennicutt, R. 1987, \aaps\ 69, 311
%
\bibitem[Walterbos \& Kennicutt(1988)]{walterbos88}
        Walterbos, R., Kennicutt, R. 1988, \aap\ 198, 61
%
\bibitem[Walterbos \& Braun(1994)]{walterbos94}
        Walterbos, R., Braun, R. 1994, \apj\ 431, 156 
%
\bibitem[Wilkinson et al.(2004)]{wilkinson}
        Wilkinson, M., Kleyna, J., Evans, N., Gilmore, G., Irwin, M.,
        Grebel, E. 2004, \apj\ 611, 21
%
\bibitem[Yoachim \& Dalcanton(2005)]{yoachim}
        Yoachim, P., Dalcanton, J. 2005, astro-ph/0501394
%
\bibitem[Zucker et al.(2004)]{zucker}
        Zucker, D. et al. 2004, \apj\ 612, 117L
%
\end{thebibliography}
\end{document}